# The Homogeneity of Interstellar Elemental Abundances in the Galactic Disk[1]

Stefan I. B. Cartledge

*Department of Physics and Astronomy, Louisiana State University, Baton Rouge, LA 70803*

scartled@lsu.edu

J. T. Lauroesch, David M. Meyer

*Department of Physics and Astronomy, Northwestern University, Evanston, IL 60208*

jtl@elvis.astro.nwu.edu, davemeyer@northwestern.edu

and

Ulysses J. Sofia

*Department of Astronomy, Whitman College, Walla Walla, WA 99362*

sofiauj@whitman.edu

## ABSTRACT

We present interstellar elemental abundance measurements derived from Space Telescope Imaging Spectrograph echelle observations of 47 sight lines extending up to 6.5 kpc through the Galactic disk. These paths probe a variety of interstellar environments, covering ranges of nearly four orders of magnitude in molecular hydrogen fraction $f(H_2)$ and more than two in mean hydrogen sight line density $\langle n_{\rm H} \rangle$. Coupling the current data with Goddard High Resolution Spectrograph data from 17 additional sight lines and the corresponding *Far Ultraviolet Spectroscopic Explorer* and *Copernicus* observations of $H_2$ absorption features, we explore magnesium, phosphorus, manganese, nickel, copper, and germanium gas-phase abundance variations as a function of $\langle n_{\rm H} \rangle$: density-dependent depletion is noted for each element, consistent with a smooth transition between two abundance plateaus identified with warm and cold neutral interstellar medium depletion levels. The observed scatter with respect to an analytic description of these transitions implies that total elemental abundances are homogeneous on length scales of hundreds of parsecs, to the limits of abundance measurement uncertainty. The probable upper limit we determine for intrinsic variability at any $\langle n_{\rm H} \rangle$ is 0.04 dex, aside from an apparent 0.10 dex deficit in copper (and oxygen) abundances within 800 pc of the Sun. Magnesium dust abundances are shown to scale with the amount of silicon in dust and, in combination with a similar relationship between iron and silicon, these data appear to favor the young F and G star values of Sofia & Meyer (2001; ApJ 554, L221) as an elemental abundance standard for the Galaxy.

## 1. Introduction

Studies of interstellar elemental abundances provide important details about environmental conditions in the interstellar medium (ISM), the composition of both its gas and dust phases, and Galactic chemical evolution. Observations of distant O and B stars in the UV are particularly useful for measuring the gas-phase abundances because of the large volume of the galaxy that these UV-bright stars open up to investigation and the variety of resonance lines for diverse elements at these wavelengths. As yet, however, large databases of such measurements are in short supply, since only a limited number of UV instruments with sufficient sensitivity and spectral resolution to isolate individual interstellar absorption components have been launched. Moreover, such facilities have recently been part of multi-purpose observatories, so that the high demand for observing time has constrained efforts to accumulate a large number of uniformly-made high-quality abundance measurements. In this paper, we present a compilation of magnesium, phosphorus, manganese, nickel, copper, and germanium UV absorption-line profiles from three recent *Hubble Space Telescope (HST)* Space Telescope Imaging Spectrograph (STIS) observing programs, including two SNAPSHOT surveys designed to study interstellar paths with a minimum of bias and a separate study of translucent sight lines.

This database has been compiled for the purpose of characterizing interstellar elemental abundance fluctuations using the latest high quality data from *HST* and *Far Ultraviolet Spec-*

---

[1]Based on observations with the NASA/ESA *Hubble Space Telescope (HST)* and the NASA-CNES-CSA *Far-Ultraviolet Spectroscopic Explorer (FUSE)*. *HST* spectra were obtained at the Space Telescope Science Institute, which is operated by the Association of Universities for Research in Astronomy, Inc. under NASA contract No. NAS 5-26555; *FUSE* is operated for NASA by the Johns Hopkins University under NASA contract NAS-32985.

*troscopic Explorer (FUSE)*. While earlier instruments such as *Copernicus* and *International Ultraviolet Explorer (IUE)* laid the foundation for understanding elemental abundance variation in the ISM (Bohlin et al. 1983; Jenkins et al. 1986; Van Steenberg & Shull 1988), the effectiveness with which abundances could be determined using each of these instruments was hampered by limitations in dynamic range and spectral resolution (*IUE*) or sensitivity and simultaneous spectral coverage (*Copernicus*). Consequently, although abundances derived from data acquired with these instruments have been useful in distinguishing the different levels of depletion corresponding to diverse interstellar environments, large uncertainties associated with unresolved cloud components and the limited number of sight lines observed with resolved features have constrained their utility for probing ISM homogeneity (sight line to sight line abundance variations). Within the last dozen years, the most productive UV instruments have been Goddard High-Resolution Spectrograph (GHRS), STIS and *FUSE*.

Interstellar abundance measurements were a specialty of GHRS. The GHRS detector's broad dynamic range and high spectral resolution in the echelle observing mode allowed very high S/N values ( $\gtrsim$ 1000; Meyer et al. 1994) to be achieved. Detailed measurements of elemental depletions in individual absorption components along several sight lines (e.g., $\zeta$ Oph, $\xi$ Per, $\mu$ Col) have significantly added to our knowledge of the different ISM phases; however, GHRS' narrow spectral windows (5.5–9 and 8.5–17 Å in the echelle A and B modes) and the heavy demand on non-GHRS *HST* observing time limited the instrument's capacity for undertaking surveys of interstellar abundance, especially in multiple elements. Nevertheless, in combination with molecular hydrogen column densities derived from *Copernicus* spectra, GHRS data have suggested that abundant elements such as carbon, nitrogen, and oxygen, as well as relatively rare but undepleted krypton, are homogeneously distributed in the Galactic disk, at least within 500 pc of the Sun (Cardelli et al. 1996; Cardelli & Meyer 1997; Meyer, Cardelli, & Sofia 1997; Meyer, Jura, & Cardelli 1998). The broader bandpasses available through STIS ($>$ 200 Å in the high-resolution $R \approx 114,000$ echelle modes) made this successor to GHRS an ideal instrument for studying interstellar abundances, since absorption features from a variety of elements are present in the same spectral window. Supplementing GHRS data with oxygen and krypton abundance measurements from the STIS programs used by this paper have strengthened the conclusion that these elements are distributed homoegenously in the ISM (Cartledge et al. 2001, 2003, 2004), while suggesting that depletion along a given path may accurately be expressed as a function of mean sight line density $\langle n_{\rm H} \rangle$ convolved with measurement uncertainty. To test these implications for their applicability to the ISM in general, however, it must be demonstrated that they are consistent with results derived from other elements.

In this paper, the database of sight lines we accumulated in studying krypton and oxygen abundance variations (see § 2) is plumbed for magnesium, phosphorus, manganese, nickel, copper, and germanium absorption features. Adopting the molecular hydrogen column densities previously determined for each sight line from *FUSE* data, the degree of variation in the gas-phase abundance ratio to hydrogen for each of these elements is examined. Depletion signatures are derived as a function of mean hydrogen sight line density in § 3, from abundance data for 47 STIS sight lines and 17 GHRS paths extending through the Galactic disk. The implications for dust phase abundances are discussed in § 4.

## 2. Observations and Data Extraction

The STIS observations of Mg II ($\lambda$1239,1240), P II ($\lambda$1302), Mn II ($\lambda$1197), Ni II ($\lambda$1317), Cu II ($\lambda$1358), and Ge II ($\lambda$1237) that are presented in this paper were acquired in the course of three *Hubble Space Telescope (HST)* programs (GO8241, GO8273, GO8662) targeting Galactic O- and B-type stars. Each exposure was made using one of three STIS setups: the $0.2''\times0.2''$aperture and the E140H grating centered at 1271 Å, the same grating exposed through the $0.2''\times0.09''$aperture, or the first aperture with the E140M grating centered at 1425 Å. The data from each exposure were then calibrated using the STSDAS[2] STIS data reduction package, although the final spectral extraction steps for observations acquired with medium and high spectral resolution differed. The IDL[3] routine STISEXTRACT (Howk & Sembach 2000) was developed in order to reliably estimate the on-order flux background in high-resolution ($R \approx 114,000$) STIS echelle spectra; experimentation has shown that improvements to the CALSTIS spectral extraction algorithm included with STSDAS now reliably reproduce the central zero-flux level for saturated absorption profiles in the output spectra. However, in the interest of consistency with procedures adopted by Cartledge et al. (2001, 2003, 2004) in publishing the krypton and oxygen abundances for these sight lines, STISEXTRACT was used for each E140H observation. The standard CALSTIS extraction algorithm was used for each medium resolution ($R \approx 45,800$) exposure. The diversity of absorption profile shapes in this database is illustrated in Figure 1, where examples of Ge II $\lambda$1237 features are plotted. *FUSE* data were also used in this analysis. Molecular hydrogen column densities were derived from these spectra, so that elemental gas-phase abundances could be compared with the total hydrogen abundance, $N(\mathrm{H}_{total}) = N(\mathrm{H\ I}) + 2N(\mathrm{H}_2)$, along each sight line. The standard *FUSE* reduction software CALFUSE, in nearly all cases v1.8.7, was applied to each raw dataset to produce a calibrated spectrum. The *FUSE* observations

---

[2]STSDAS is a product of the Space Telescope Science Institute, which is operated by AURA for NASA.

[3]IDL is an acronym for Interactive Data Language, a common programming tool developed by Research Systems, Inc.

used in this analysis are listed along with the STIS datasets in Table 1.

Following the procedures outlined in previous papers (Cartledge et al. 2001, 2004), column densities for each element in the spectra for each sight line were determined independently using two methods: apparent optical depth analysis (Savage & Sembach 1991; Jenkins 1996) and profile fitting (Welty, Hobbs, & York 1991; Mar & Bailey 1995). The $f$-value adopted for each absorption line studied is listed in Table 2. The measurements associated with each method and the column density adopted in consequence for each sight line are presented in Table 3. Only abundance measurements considered to be reliable are included in this table—dubious results, non-detections, and severely blended profiles being omitted. To evaluate the accuracy of these results in terms of unresolved saturation, we refer to simulations by Savage & Sembach (1991) of apparent optical depth measurement reliability; for ease of discussion, the expression $|N_A(x) - N_P(x)|$ is adopted for the difference between column densities derived using our two methods, expressed in logarithmic form. In order to simulate real spectra, Savage & Sembach (1991) constructed a variety of idealized absorption profiles and convoluted them with instrumental response functions. Column densities were derived for the resulting features using both optical depth and curve of growth methods, assuming a series of $f$-values. Based on this framework of measurements, the accuracy of a given apparent optical depth measurement can be estimated.

Four lines in particular were used as guides for estimating the influence of unresolved saturation on the features we studied: P II $\lambda$1302 (the strongest line for which we publish column densities), Ge II, and the Mg II $\lambda$1239/1240 doublet. For phosphorus and germanium, apparent optical depth column densities generally agreed within error with the results derived through profile-fitting. Unfortunately, these features were not present in all sight lines. The magnesium doublet was much more useful, both because it was present along more sight lines and because comparison of column densities derived from apparent optical depth for two lines with different $f$-values is a more sensitive gauge of unresolved saturation. The Savage & Sembach (1991) simulation framework suggests that if lines whose $f$-values differ by a significant factor lead to column density profiles in velocity space that are roughly equivalent, then it is unlikely that unresolved saturation is a serious concern. Alternatively, if the stronger line indicates a measurably smaller value for $N$, the unsaturated result can be inferred from the size of the gap and the larger abundance.

In this collection of lines, the magnesium doublet possessed an appropriate $f$-value ratio for flagging the presence of *unresolved* saturation; for reference, it should be noted that absorption associated with these lines often reached depths of 80% and 60% of the continuum for Mg II $\lambda$1239 and Mg II $\lambda$1240, respectively. Despite the strength of this absorption, apparent-optical-depth column densities determined from each feature agreed with their counterparts within their respective statistical uncertainties (at most 0.04 dex) for the vast majority of the database sight lines. The few paths with strong narrow absorption features were the only exceptions, exhibiting differences of up to 0.08±0.04 dex, generally, and 0.18±0.04 dex for the sight line toward HD147888. The larger of the values for the 1239 and 1240 Å lines has been listed in Table 3 as representative of the apparent-optical-depth column density for each sight line. According to Savage & Sembach (1991), such column density discrepancies are symptomatic of unresolved saturation. Their prescribed correction procedure, valid for column density gaps of up to 0.10 dex in isolated Gaussian profiles for lines whose $f\lambda$ products differed by a factor of 2, devolved to simply adding the difference in column densities to the higher value, given an instrumental spectral resolution similar to the STIS high-resolution echelle mode. For the current sight lines, applying this admittedly simplistic correction produced column densities that agreed within error with the results determined by profile fitting. In the most severe case, HD147888, the additive logarithmic correction for a discrepancy of 0.18 dex is 0.27 dex, but in light of the uncertainty in the gap between $\lambda$1239 and $\lambda$1240 column density measurements, the profile-fitting result is in reasonable agreement with the corrected apparent-optical-depth value ($|N_A(\mathrm{Mg}) - N_P(\mathrm{Mg})|_{\mathrm{HD147888}} = 0.19$ dex). Thus, since profile-fitting column density measurements are consistent with "corrected" apparent-optical-depth results, they are adopted for all sight lines as reasonable determinations of the full sight line magnesium column density.

One or two notes of caution are warranted, however, before further analysis is made. Jenkins (1996) re-examined the apparent optical depth technique in detail, deriving an alternate correction procedure to the Savage & Sembach (1991) method. It was found that the latter algorithm worked well for absorption profiles resembling a gaussian, but suffered relative to the alternate procedure for a broader range of profiles which are perhaps more realistic. Nevertheless, in cases with maximum depths similar to those in the present sample, the corrections are approximately the same. A second point of consideration arises out of a recent study of oxygen absorption in a smaller set of sight lines that nevertheless overlaps our sample to some extent. Jensen, Rachford, & Snow (2005) included equivalent width measurements from oxygen lines in *FUSE* and *IUE* spectra with their STIS results in order to determine column densities using curves of growth. Unfortunately, the resulting curves were often based on equivalent widths from only a few lines; nevertheless, Jensen, Rachford, & Snow (2005) suggest that for one of the four cases in which there were notable discrepancies from previous work, there is evidence that profile-fitting may have missed some saturation. However, if the profile-fitting technique used in this paper does not entirely account for line-of-sight elemental abundances, it is less likely to affect the shape of the observed dependence of column density (or depletion) on mean sight line density than generate an additional source of scatter on top of the overall trend.

Since $|N_A(x) - N_P(x)|$ values for other elements discussed in this paper are all consistent with the information gleaned from the phosphorus, germanium, and magnesium absorption-profile examinations, it is concluded that unresolved saturation is not of serious concern for these data, except along sight lines with strong narrow absorption profiles. Using magnesium, it has been demonstrated that profile-fitting column densities recover, within error, the values that would be derived by adjusting apparent-optical-depth determinations according to the Savage & Sembach (1991) algorithm. Hence, the profile-fitting column densities have been adopted for all elements for each sight line. For HD147888, this procedure would appear, at worst, to underestimate the true column density by 0.08 dex; however, measurement uncertainties for all other dataset sight lines are larger than any uncertainty introduced by the adoption of profile-fitting results. Column density errors for each sight line have been assigned according to the prescription specified for krypton and oxygen results, namely, the quadrature sum of absolute errors for continuum placement (the listed apparent-optical-depth uncertainty) and profile-fitting. Unless otherwise stated, it should be assumed that all errors in the tables, figures, and text are 1-$\sigma$ uncertainties.

While individual orders of the STIS spectra provided the elemental abundance information used in this paper, in combination they also shed light on the atomic hydrogen column density in the form of the Lyman-$\alpha$ line. Using a continuum-reconstruction algorithm (Bohlin 1975; Diplas & Savage 1994), $N$(H I) was derived for each sight line not subject to serious stellar Lyman-$\alpha$ contamination. Molecular hydrogen column densities were determined from *FUSE* spectra by fitting absorption profiles for $J$=0 and 1 rovibrational transitions, where observations were available. In the absence of a reliable measurement of either H I or $H_2$, the hydrogen column density was inferred from the krypton and/or oxygen abundance (see § 3). Since the current sight line selection was based on the detection of O I $\lambda$1355 absorption, details of the values adopted for both H I and $H_2$ column densities may be found in Cartledge et al. (2004). The total hydrogen abundance for each sight line is listed along with the corresponding elemental abundances in Table 4.

## 3. Gas-phase Abundance Variability

In a previous discussion of interstellar oxygen abundances (Cartledge et al. 2004), the 4-parameter Boltzmann function was very successful in resolving scatter in the gas-phase O/H abundance ratio into a dependence of O/H on $\langle n_{\rm H} \rangle$. Prior studies had generally indicated that this ratio was constant within 500 pc of the Sun—mirroring, in this respect, the behavior of proportions relating the abundance of the elements carbon, nitrogen, and krypton to hydrogen. Spectra probing translucent sight lines, however, suggested that oxygen depletion was somewhat enhanced in this group, although krypton abundances remained consistent with levels evident for more rarified interstellar paths (Cartledge et al. 2001). Supplementing the earlier STIS sample with more than 20 additional sight lines, it became apparent that although the depletion enhancement was of the same order of magnitude as the measurement uncertainties, it nevertheless distinguished two ranges of mean sight line density with roughly constant O/H ratios. The function that describes the transition between such plateaus has a Boltzmann form:

$$(x/\mathrm{H})_{gas} = (x/\mathrm{H})_c + \frac{(x/\mathrm{H})_w - (x/\mathrm{H})_c}{1 + e^{(\langle n_\mathrm{H}\rangle - \langle n_0\rangle)/m_0}}, \quad (1)$$

where $(x/\mathrm{H})_w$ and $(x/\mathrm{H})_c$ are the gas-phase abundance levels for the element $x$ that are appropriate for low and high mean sight line densities, respectively, $\langle n_0\rangle$ is the inflection point of the curve, and $m_0$ is its slope at the inflection point. Its success in distinguishing between subtly different degrees of oxygen depletion makes this function appropriate for the current analysis, as a guide to the interpretation of gas-phase elemental abundance trends. Elemental abundances for individual sight lines are not directly comparable with this fitting function without normalizing the data to the path's total hydrogen abundance. Unfortunately, the observed star was not of a type appropriate for direct hydrogen column density determinations in some cases, or in others the data were simply not available. In order to derive the fullest use from our database, seven sight lines required that the hydrogen column density be estimated through other means than direct measurement.

Recently, Cartledge et al. (2003) derived a value of $\log_{10}(\mathrm{Kr/H}) = -9.02\pm0.02$ for the interstellar Kr/H abundance ratio, where the data-point dispersion was less than column-density uncertainties (0.08 dex). Consequently, since krypton column densities have been determined for the HD43818, HD148594, and HD175360 sight lines[4], they have been assigned $N(\mathrm{H})$ and $\langle n_\mathrm{H}\rangle$ values in Table 4; their associated uncertainties were derived from the quadrature sum of Kr/H scatter and krypton measurement uncertainties. The resulting mean sight line density for each path is greater than 1 cm$^{-3}$; in particular, the value for HD148594 is $\log_{10}\langle n_\mathrm{H}\rangle = 0.72$, comparable to figures associated with "translucent" ISM sight lines such as HD27778, HD37021, and HD37061. It should be noted that in using the word translucent here, we refer not to individual clouds but to sight lines whose overall properties are consistent with the van Dishoeck & Black (1989) definition; Rachford et al. (2002) studied several sight lines with $A_V \gtrsim 1$ mag, including four of our paths, and found none likely to intersect a translucent cloud.

---

[4]The HD43818 measurement [$\log_{10}N(\mathrm{Kr}) = 12.69\pm0.09$] is somewhat poorer in quality than the other two measurements; consequently, the value has only previously been published parenthetically (Cartledge et al. 2004).

The Boltzmann function defined by Cartledge et al. (2004) parameterized a relationship between the O/H abundance ratio and $\langle n_{\rm H} \rangle$ that reduced scatter to just slightly larger than observational uncertainty. Moreover, the transition region between the two plateaus isolated by fitting this function to the oxygen data encompasses only a very small range of $\langle n_{\rm H} \rangle$. Thus, in the absence of direct hydrogen or krypton measurements, it is reasonable to assign hydrogen column densities using $N$(O) and one of the fitted (O/H)$_w$ or (O/H)$_c$ parameters, provided that the sight lines can be appropriately classified by mean sight line density. Since the difference between these parameters (0.14 dex) is significant relative to the sum of observational and measurement errors, special care was taken to ensure that this classification was justified. Applying $N$(O) to (O/H)$_w$ and (O/H)$_c$ gave two candidate hydrogen abundances for each sight line in question. Ratios of the various elemental column densities to each of these hydrogen values were compared with results characteristic of the corresponding density regime and the better match used to identify which value was adopted for the path. This procedure may introduce a slight bias to $\langle n_{\rm H} \rangle$ plots that appear later in this section, but it is reassuring that regardless of which of the hydrogen abundances was chosen, sight lines toward HD52266, HD71634, HD111934, HD156110 were clearly identified as low density paths. The nature of HD36841 measurements is more ambiguous, since calibration using $N$(O) suggests that the mean hydrogen sight line density is between 1.0 and 1.4 cm$^{-3}$, in the middle of the transition region between the low- and high-$\langle n_{\rm H} \rangle$ plateaus (Cartledge et al. 2004). In this case, we adopt a value for $N({\rm H})_{\rm HD36841}$ consistent with the high-density oxygen plateau. Errors associated with all the oxygen-based calibrations were derived from the quadrature sum of oxygen abundance error and the O/H ratio dispersion (0.09 dex).

### 3.1. Lightly-Depleted Phosphorus and Germanium

The elements included in the current database are depleted from the gas phase to a variety of degrees. Because there is exchange of material between the different phases of the ISM, the intrinsic level of variation in a given element's gas-phase abundance will in part reflect the degree to which it is depleted generally. Specifically, for an element that exists primarily in a gaseous state, abundance fluctuations in an absolute sense should approximate the variability of its total interstellar abundance. Similarly, a heavily depleted element would exhibit a very large scatter in gas-phase measurements even if fluctuations in the total (or dust phase) abundance are small. To set the stage for gauging overall interstellar abundance variability, we start with lightly depleted elements.

The variation of gas-phase interstellar abundance on mean hydrogen sight line density has previously been discussed using the current dataset for elements either that are un- or only lightly depleted (Kr and O, respectively; Cartledge et al. 2003, 2004). The next most lightly depleted elements in the current suite are phosphorus and germanium—abundance ratios relative to hydrogen for these elements are compiled in Table 4 and presented in Figure 2. Phosphorus is produced primarily as a result of neon burning and explosive nucleosynthesis (Anders & Grevesse 1989); germanium is formed by both the *r*- and weak *s*-processes, with further contributions from the *e*-process (nuclear statistical equilibrium; Anders & Grevesse 1989). In situations where phosphorus is depleted from the gas phase, it is expected to appear in the form of PN and some PO in molecules or as phosphine ($PH_3$) on grains in dense clouds (Charnley & Millar 1994). The dominant form for germanium depleted from the gas phase in the diffuse ISM, or even meteorites, is not known (Greshake et al. 1998).

It is clear in Figure 2 that the depletion level for each of these elements undergoes a significant adjustment as mean sight line density increases. In warm diffuse clouds, the current gas-phase abundances for phosphorus and germanium approximately match the levels estimated by Welty et al. (1999) from a combination of previously-published *Copernicus* and GHRS measurements for individual components. And notably, the Boltzmann functions describing the data have roughly the same inflection point as the function previously published for oxygen abundances determined from the same sight lines (see Table 5). At mean hydrogen sight line densities higher than these values, phosphorus and germanium column densities ratios to hydrogen plateau at values somewhat lower than the levels estimated by Welty et al. (1999). The only sight line that appears to stand distinct from the scatter expected from measurement uncertainty in each case is HD156110.

The discrepancy between HD156110 and the bulk of the current dataset is perhaps more easily seen in Figure 3, which depicts the scatter in phosphorus, germanium, oxygen and krypton as a function of $\langle n_{\rm H} \rangle$. The shaded regions in each panel indicate the bounds within which approximately $\frac{2}{3}$ of the data should appear, given the size of a typical measurement uncertainty. The bold error bar at $\log_{10}\langle n_{\rm H} \rangle = 0.90$ indicates the corresponding size of the measured scatter, which in each case is either consistent with or smaller than the width of the expected statistical scatter. The breadth of the phosphorus data distribution is influenced both by HD156110 and HD36841. The latter sight line has a large uncertainty in its length (up to 25%) derived from an ambiguity with which group of stars it is physically associated; taking into account the proximity of its specific mean sight line density to the inflection point of the fitted Boltzmann curve implies that any marked deviation may be in part due to an error in pathlength. The hydrogen column density toward HD156110 was calibrated using the oxygen abundance determined from O I] $\lambda 1355$, since its Ly-$\alpha$ profile is an intractable blend of similarly strong stellar and interstellar absorption components. It is also possible that this path has a much larger effective $\langle n_{\rm H} \rangle$-value than is apparent, since the star resides in the low halo and much of the intervening space may be empty. In fitting the data for each element with a Boltzmann function, a few sight lines were pruned from the input set in order to derive a stable representative solution. In particular, HD156110 and HD36841 were removed from consideration for phosphorus and HD156110 alone for germanium. Determination of the scatter listed in Table 5, however, included all data points for each element.

### 3.2. Moderate and Heavily-Depleted Elements

The goal of evaluating elemental abundance variability for more heavily depleted elements is complicated by the small proportion of the total amount of each element that exists in the gas phase. For example, Welty et al. (1999) estimate that in warm neutral clouds only 25% of magnesium would be in the form of atomic gas. Thus, if variations in the dust abundance making up the bulk of interstellar magnesium are only of order 5% of its mean value, the least variation to be expected in warm-ISM gas-phase values would be 15%. This argument assumes that the total elemental abundance, relative to hydrogen, is absolutely uniform, or in other words that any deviation in the dust abundance is matched by an equal and opposite fluctuation in the gas amount. A more realistic treatment would allow only partial anti-correlation between gas and dust abundances, synonymous with some intrinsic level of variation in the total elemental abundance ratio to hydrogen. Since we are investigating fluctuations relative to a Boltzmann function description depending on mean hydrogen sight line density, a base level of fluctuation *must* be present due to the variety of sight line types that will result in a single numerical value of $\langle n_{\rm H} \rangle$ (e.g., a rarified interstellar path intersecting a dense cloud can yield the same value as a path of uniform moderate density). The effect of increased scatter in gas-phase abundance for heavily depleted elements is most easily discerned by comparing the scatter in copper measurements in Figures 4 and 5 with germanium, since the gap between warm and cold ISM depletion levels for these two elements expressed in logarithmic form is similar.

The elements magnesium, manganese, copper, and nickel, whose abundance distributions are depicted in Figures 4 and 5 and whose Boltzmann parameter fits are listed in Table 6, are produced at various stages of massive star evolution (Anders & Grevesse 1989). These elements are presented in each plot in order of increasing warm ISM depletion, with magnesium and nickel having the most and least fractions, respectively, among them in the gas phase. Nevertheless, because magnesium is also by far the most abundant element in this group, it is also the most significant contributor by mass to dust. In particular, it is a common constituent in silicate grains (e.g., $Mg_2SiO_4$, Mg[Ca,Fe]$SiO_4$, $MgSiO_3$; Draine 2004; Kemper, Vriend, & Tielens 2004) used in dust models to account for observed infrared absorption from Si-O stretching and O-Si-O bending modes. Because the other elements in this group, manganese, copper, and nickel, are so comparatively rare in the ISM, the forms they take upon depletion from the gas phase in the diffuse medium are infrequently discussed in the literature. In meteorites, however, these elements have been found in at least metallic, sulfide, and oxide forms (e.g., Petaev et al. 1992; Kleinschrot & Okrusch 1999; Jones, Grossman, & Rubin 2005).

A common feature of the elemental abundance distributions in Figure 4 is the relative shallowness of the slope in Boltzmann functions fit to these data when compared with the fits to phosphorus, germanium, and oxygen. This is coupled with a shift in the transition midpoint to smaller, even unphysical, values. In particular, the fitted $\langle n_0 \rangle$ parameters for magnesium and nickel are negative, apparently in consequence of the need to account for enhanced scatter in the data for mean hydrogen sight line densities roughly between 0.20 and 2.0 cm$^{-3}$. These cases identify a complication inherent in using $\langle n_{\rm H} \rangle$ to gauge elemental abundance—namely, this sight line property has an uncertainty not only associated with the hydrogen column density measurement but also with the adopted pathlength. In detail, an underestimated column density would shift a datapoint diagonally to lower mean density and higher abundance ratio to hydrogen, spreading the data roughly parallel to the slope of the Boltzmann function in the transition region. Misjudging the length of a sight line, however, distributes the data more broadly off the midline of the Boltzmann function through the transition region. This effect is perhaps most clearly seen in considering the position of HD36841 in the top panel of Figure 2. This sight line exhibits gas-phase abundances consistent with those of higher mean density paths for each detected element. HD36841 has been grouped with HD37903 in Orion OB1b by Brown, de Geus, & de Zeeuw (1994) with a mean distance of 360±70 pc, since these authors did not find the distinctions that led Warren & Hesser (1978) to subdivide this group. However, if HD36841 is in a subgroup 20% closer than that including HD37903, as indicated by Warren & Hesser (1978), then it loses the status of extreme outlier in Figures 3 and 5 as it conforms more closely to the current Boltzmann fits. In all cases, the pathlength adopted in deriving a value for $\langle n_{\rm H} \rangle$ is either the distance to the cluster of which the star in question is a member or, where this is not available, its *Hipparcos*-based and/or spectroscopic parallax. We consider the error for HD36841 to be unusually large relative to other datapoints in the transition regions in these plots. For sight lines away from such regions, distance uncertainty will add only a very small contribution to the scatter, as the data settle into plateaus.

The difficulties introduced by any error in distance couple with measurement uncertainties, the increased scatter in gas abundances expected for moderate to heavily depleted elements, and the larger gaps between warm and cold neutral ISM depletion levels to make it more difficult to derive a stable solution with a realistic transition point. So to investigate elemental abundance homogeneity we have adopted the non-physical two-plateau solutions for magnesium and nickel in Table 6, since they represent the data fairly well in the mean density interval in question despite larger $\chi^2_\nu$ values. Prior to proceeding with the scatter analysis, however, it is appropriate to explore a tantalizing explanation for the difficulty encountered fitting our second selection of elements.

In an idealized picture of abundance variations as a function of mean hydrogen sight line density, Spitzer (1985) calculated that if canonical depletion levels were adopted for warm neutral gas and diffuse and large cold clouds, these being distinguished by their extinction, size, and spatial frequency properties, then each would dominate the distribution of abundances over a range in $\langle n_{\rm H} \rangle$. The ranges that were determined for each "cloud" type were less than 0.2 $\rm cm^{-3}$, between 0.2 and 3.0 $\rm cm^{-3}$, and greater than 3.0 $\rm cm^{-3}$, respectively. Coincidentally, the onset of increased scatter in the magnesium, manganese, and nickel distributions in Figure 4 occurs near $\langle n_{\rm H} \rangle = 0.2$ $\rm cm^{-3}$ and it diminishes near 2.0 $\rm cm^{-3}$ ($\log_{10}\langle n_{\rm H} \rangle$ $= -0.70$ and 0.30, respectively), where the latter value is within the gap between the transition midpoint for lightly depleted elements and the mean density supposedly distinguishing diffuse and large cold clouds. We investigated this idealized picture by fitting a three-plateau Boltzmann function to the magnesium and nickel data; the results are presented in Figure 6 and Table 6. The three-plateau Boltzmann functions qualitatively improve the fit, adhering more closely to the distribution of the elemental abundance data. However, neither the quality of fit, as expressed by $\chi^2_\nu$, nor the scatter with respect to the function benefit from adding a third mean density range of near-constant depletion. Nevertheless, the warm ISM abundance plateaus from the three-plateau fits may be adopted for magnesium and nickel as references, since these are consistent with the data at the low density end of the $\langle n_{\rm H} \rangle$ range we have investigated.

A further notable characteristic of the nickel distribution is the much lower abundance for four datapoints than is typical of other sight lines with similar $\langle n_{\rm H} \rangle$ values. Among the other elements we have studied for this project, gas-phase abundances for these sight lines are not unusual for paths with similar mean hydrogen densities. Nor is there evidence for nickel being in either different ionization or excitation states. Within this group of four, HD27778 has also been observed at wavelengths that include weak carbon, silicon, and iron absorption features (Sofia et al. 2004; Miller et al. 2005). These elements reveal a sight line with unusually strong depletion: C II is undetected in absorption (a $3\sigma$ upper limit on the gas-phase carbon abundance leaves it four standard deviations below the mean diffuse ISM C/H ratio) and the gas-phase Si/H and Fe/H ratios are roughly one half and one fifth of values typical for paths with similar $\langle n_{\rm H} \rangle$-values. Unfortunately, this sub-branch of the nickel abundance pattern, however, is not readily explained by the link Miller et al. (2005) identified between grain size and silicon and iron depletion. Those authors noted that among their sample of high mean density sight lines, which is a subgroup of the current database, smaller values of $R_V$ or the Fitzpatrick & Massa (1990) fitting parameter $c_4$ implied heavier iron and silicon depletion, whereas such values for HD37903 and HD203532 are larger than average (Valencic et al. 2004).

## 3.3. Variation with Other Sight Line Properties

### *3.3.1. Spatial Abundance Fluctuations*

Before discussing the implications of the facility with which gas-phase elemental abundances are described by the 2-plateau functions we have adopted, dependencies on other sight line properties should be examined so that their influence on scatter may be gauged. One notable trend previously identified in oxygen measurements is a relatively low mean abundance for low density ($\langle n_{\rm H}\rangle < 1.0$ cm$^{-3}$ or $\log_{10}\langle n_{\rm H}\rangle < 0.0$) short sight lines when compared with values typical of longer paths (André et al. 2003; Cartledge et al. 2004).

As shown in Figure 7, copper joins oxygen in exhibiting a lower mean ratio to hydrogen for low density sight lines shorter than 800 pc. Moreover, the magnitude of the abundance gap between short and long paths is 0.129±0.033 dex in copper, a value consistent with the 0.099±0.028 difference measured for oxygen. The simplest physical explanation for the existence of such a distinction between local and long distance elemental abundances would be infall of low metallicity gas onto the Galactic disk near the Sun, diluting the local ISM (Meyer et al. 1994). A key feature of this scenario is an identical gap for all elements, which suggestively is consistent with the results for copper and oxygen. It should also be noted that when low-density oxygen measurements are separated into the local and distant samples, the scatter in each group is reduced and, in particular, the local scatter (see Table 7) mirrors the breadth of the krypton abundance distribution in Table 5. The sight lines with krypton measurements are predominantly shorter than the 800 pc dividing line we have noted and despite also including dense sight lines, it is not unreasonable to compare these scatter values because krypton is undepleted. Unfortunately, these points are only suggestive, and since the current STIS database does not include many short sight lines and GHRS measurements of our other elements have not been published for the sight lines we include in Figure 7, the effect cannot be assessed more rigorously with these data. Many more high quality abundance measurements in a variety of elements are needed for that purpose. With respect to other spatial sight line characteristics (e.g., sky position), no significant trends have been identified in the full database.

### *3.3.2. $E(B-V)$, $E(B-V)/d_*$, and $f(H_2)$*

Sight line properties other than mean hydrogen sight line density have been used to distinguish classes of elemental depletion with various degrees of success. Jenkins et al. (1986), Van Steenberg & Shull (1988), and Snow, Rachford, & Figoski (2002), in particular, investigated fluctuations in gas phase abundance with a variety of diagnostics including several color excesses, both singularly and scaled by length of sight line, extinction parameters related to the UV bump, gas-to-dust ratios, and molecular hydrogen fraction, $f(\mathrm{H}_2)$. The sight line property $\langle n_{\mathrm{H}} \rangle$ performed best as a discriminant of distinct elemental depletion levels despite its inexact reflection of interstellar conditions and the large scatter apparent in the abundance distributions. The latest generation of ultraviolet spectrometers has improved the quality of abundance measurements to the point where a simple functional dependence is clearly evident with only a small level of scatter, so the improvements in measurement quality might reveal subtle trends with respect to properties other than $\langle n_{\mathrm{H}} \rangle$ that were hitherto masked by error considerations. In search of such trends, plots of abundance as a function of other sight line characteristics have been constructed; since we are studying elements with different intrinsic levels of depletion, they are shown for both germanium (Figure 8) and magnesium (Figure 9).

The quantity $f(\mathrm{H}_2)$ has been a popular choice for classifying sight lines by their environmental conditions because of its physical significance. By definition, $f(\mathrm{H}_2)$ expresses the proportion of hydrogen along a given path that is in molecular form; thus, it would seem to be ideal for distinguishing the elemental depletion levels associated with different environments—namely, diffuse gas and dense clouds. But in practice, the upper right panels of Figures 8 and 9 imply that $\mathrm{H}_2$ exists more ubiquitously than was imagined before the *FUSE* translucent cloud $\mathrm{H}_2$ survey ascribed the bulk of their selected sight lines to the overlap of multiple diffuse clouds rather than denser structures (Rachford et al. 2002).

The $B-V$ color excess is a rough measure of the amount of dust in a sight line, so one might expect gas-phase elemental abundances to be related to this property. This quantity is cumulative, however, and since paths with a wide range of lengths are included in this database, depletion levels associated with different interstellar environments are not distinguishable in the upper left panels of the germanium and magnesium figures. But when scaled by distance, the tight relationship between $E(B-V)$ and $N(\mathrm{H}_{total})$ in the Milky Way molds the shape of the abundance variation into a form similar to that evident using $\langle n_{\mathrm{H}} \rangle$ as the independent variable. However, when expressed with reference to $E(B-V)/d_*$, elemental abundance data are more broadly scattered (see the lower left panel of these figures), indicating that $\langle n_{\mathrm{H}} \rangle$ is the better characteristic to use in distinguishing depletion levels arising out of different interstellar environments. Nevertheless, this link between the two properties perhaps makes the clear trend with mean sight line density more satisfying on an intuitive level.

### 3.4. Direct Implications

The elemental abundance data we have introduced conform closely to the Boltzmann function when presented in terms of the mean hydrogen density along each sight line. Specifically, this implies that the depletion level for each element is constant over two ranges of physical density which are apparently associated with the warm and cold neutral ISM environments. Also, the scatter in the data relative to our adopted fitting function is consistent with measurement uncertainty for un- and lightly depleted elements and increases for those more heavily depleted. Finally, the degree of scatter in copper and oxygen data, those elements with the most low-$\langle n_{\rm H} \rangle$ measurements in the nearby ISM, appears to be enhanced by a local suppression of elemental abundances. Taken together with earlier krypton and oxygen results, these statements describing the abundance variation trends in magnesium, phosphorus, manganese, nickel, copper, and germanium lead to some interesting results.

First, the well-worn assumption that a cosmic elemental abundance reference set exists is re-affirmed by these data, which go so far as to provide a hard upper limit of 0.06 dex on intrinsic variations in total abundance relative to hydrogen based on the measured scatter in krypton data. Notably, the scatter measured among oxygen abundances for local low-density paths is also 0.06 dex. This degree of fluctuation includes measurement uncertainty, which our apparently conservative estimates typically make out to be larger than this limit. Because the error sources have been accounted for in detail, however, we do not consider the individual uncertainties to have been grossly overestimated. Nevertheless, our limit on intrinsic variability in total elemental abundance can be further refined by considering additional elements. If one assumes that gas- and dust-phase abundance fluctuations are uncorrelated, then they add in quadrature. Assuming they are also equal in magnitude, one can then use the measured scatter (Tables 5 and 6) to derive an estimate of the width of the distribution of total elemental abundances. Comparing the results with the solar photosphere abundance set, as a proxy for a cosmic standard, the widest distribution would be attributed to phosphorus with a spread of 0.069 dex. If we assume that the measurement uncertainty is of order 0.06 dex, then an allowance of approximately 0.04 dex remains for other sources of scatter, including intrinsic abundance fluctuations. Moreover, since germanium at 0.032 dex exhibits the next largest spread in total abundance according to the formula used for phosphorus, we conclude that 0.04 dex is a reasonable and more informative limit on total elemental abundance variation than the 0.06 dex limit associated with the width of the raw krypton-to-hydrogen ratio distribution. Note that these statements bear on abundance ratios to hydrogen accumulated over distances of hundreds or thousands of parsecs. Significant small-scale fluctuations due to enrichment processes, for example, are not disallowed, nor is the radial abundance gradient evident in stellar and nebular measurements contradicted.

This picture of an ISM within which elemental abundances are homogeneously mixed is consistent with other less direct indications provided separately by stellar atmosphere abundances and the analysis of meteoritic presolar dust grains. Early indications from B stars (e.g., Kilian-Montenbruck, Gehren, & Nissen 1994) and F and G stars in the Solar neighborhood (Edvardsson et al. 1993) were that elemental abundances exhibited significant intrinsic variation (up to 0.20 dex) on top of that expected from strictly measurement uncertainties. Recently, Reddy et al. (2003) analyzed a sample of stars similar to that examined by Edvardsson et al. (1993), but with data of higher quality. They reported significant scatter in stellar metallicity, but reduced scatter in elemental abundance ratios to iron relative to previous studies. Reddy et al. (2003) interpreted their broad metallicity distribution as a consequence of stellar migration and the diversity of the sample, effects which were essentially erased when the abundances of different elements were expressed as ratios to iron. Since the scatter in abundance ratios was not larger than a value attributable to measurement errors, they concluded that the Galactic thin disk is chemically homogeneous at a given epoch. Nittler (2005) reached a similar conclusion by examining presolar grains found in meteorites. Nittler (2005) argued that the comparison of measured isotopic ratios of Si and Ti in these grains with their corresponding values in Monte Carlo simulations of galactic chemical evolution place hard limits on the allowed level of chemical heterogeneity; specifically, variations of order 1–2% were inferred. It appears that these various approaches are now converging to very tightly constrain ISM chemical homogeneity, even to the level of a few percent.

In reflecting again on the current results, it must be noted that the similarities and differences between the Boltzmann parameters derived for each element are important clues to how the underlying environmental conditions in the ISM are reflected in the gas-dust-molecule balance in absorption complexes. For instance, the agreement between the $\langle n_0 \rangle$-values for phosphorus, germanium, and oxygen suggests that the reactions involved in their depletion enhancement are related; since oxygen is so abundant in the ISM, perhaps the physical density reaches a threshold beyond which this element more readily becomes involved in bonds with the other two. Because a larger proportion of magnesium, manganese, nickel, and copper are in dust, the parameters for these elements are not necessarily physical; nevertheless, there is a dramatic change in depletion levels for each of these elements, excepting nickel, that is also centered roughly where oxygen undergoes its depletion-level transition. Nickel appears either to be more gradually depleted as the environmental conditions scale from warm neutral gas to cold dense clouds or to undergo another earlier transition along this continuum. Qualitatively, it seems that the magnesium and manganese distributions may also experience such a transition, although the three-plateau fits to these elements' data are not quantitatively justified (§ 3.2). Regardless, the onset of reduced gas-phase abundances for these three elements at lower $\langle n_{\rm H} \rangle$-values than the common higher density transition implies that the average composition of dust grains changes with mean hydrogen sight line density.

## 4. Implications for Dust Composition

The tight limitations on total elemental abundance variability we have inferred imply that cosmic abundance constraints can be applied to dust models at a refined level. Although the current data place limits on fluctuations, the precise levels that should be associated with each element's abundance are still the subject of debate (Sofia & Meyer 2001).

To preface the discussion of how dust abundances fit in with the limits on gas-phase abundance identified in this paper, the adopted depletion levels for warm and cold neutral ISM environments are given in Table 8. Recently, solar reference levels for several abundant elements have undergone significant downwards revision as a consequence of improvements in solar atmosphere models (Holweger 2001; Allende Prieto, Lambert, & Asplund 2002; Asplund et al. 2004). These new standards had been shown to conflict with a large body of results from helioseismology (Basu & Antia 2004); however, more recent modelling has led to the conclusion that if the solar neon-to-oxygen abundance ratio is consistent with a seemingly constant value determined for other stars, then the latest solar abundance set, by Asplund, Grevesse, & Sauval (2005), can essentially be reconciled with helioseismology (Drake & Testa 2005). Yet Antia & Basu (2005) conjecture that adjusting the neon abundance is not likely to be the best solution: they favor also raising the abundances relative to hydrogen of elements such as carbon, oxygen, nitrogen, and iron by a few hundredths of a dex from (Asplund, Grevesse, & Sauval 2005). Consequently, we adopt Lodders (2003) solar photosphere abundances for Table 8 and subsequent discussion, since this set fits the Antia & Basu (2005) prescription and generally agrees with (Asplund, Grevesse, & Sauval 2005) within error.

The current database includes a subset of translucent sight lines that have been observed at longer wavelengths (2124–2396 Å) for the purpose of investigating the interstellar carbon abundance (Sofia et al. 2004). Weak iron and silicon features were also present in those spectra and, in combination with our magnesium abundances, these data constrain the silicate dust properties for this selection of dense sight lines. In studying the iron and silicon features, Miller et al. (2005) found that the dust-phase abundances for both increased as mean grain size decreased, where grain size was described using the sight line properties $R_V$ and $c_4$ (an extinction curve parameter; Fitzpatrick & Massa 1990). Their depiction of this trend is reproduced along with our magnesium data in Figure 10; N.B., Lodders (2003) photosphere references are used here to infer the dust abundance for each sight line. Contrary to the trend for iron and silicon, the amount of magnesium in dust appears roughly constant for these sight lines, or even to decrease if HD27778 is neglected, as more of the sight line's dust mass arises in smaller particles. Miller et al. (2005) suggested two non-exclusive rationales for the trends in iron and silicon. First, a new small grain population that takes up additional iron and silicon might be present along sight lines with lower mean grain size in addition to the grain distribution associated with the least amount of these elements in dust. Alternatively, the overall grain population has been processed to smaller sizes, opening up more surface sites for enhanced depletion. A constant magnesium dust abundance would lend support to the former explanation, while a falling trend would suggest a complicated version of the second scenario involving the preferential loss of magnesium from grains as iron and silicon accumulate. Unfortunately, the trend in magnesium is more ambiguous than that for either the iron or silicon.

Continuing the lines of investigation followed by Miller et al. (2005), Figure 11 depicts the trend in magnesium dust abundance as it relates to silicon. In plotting silicon against iron, Miller et al. (2005) noted that the data were consistent with a linear regression fit extending from translucent sight lines, where almost all of each element is in dust, to paths for which no silicon is in dust but $\mathrm{Fe/H}_{dust} \times 10^6 = 19.7$.[5] The inferred ratio of magnesium to silicon in dust is close to 1.0 for a diversity of interstellar environments; a straight line fit to magnesium data for the same sight lines and/or absorption components used by Miller et al. (2005) has the solution $(\mathrm{Mg/H})_d = (0.91{\pm}0.10){\cdot}(\mathrm{Si/H})_d + (3.4{\pm}2.7){\times}10^{-6}$. According to these fits, there are reservoirs of iron and magnesium depletion occupying approximately 20 and 3 atoms per million hydrogen that arise "before" the onset of silicon depletion into dust. Also, any additional number of iron and magnesium atoms will be depleted in a fixed proportion to the number of silicon atoms depleted, with ratio values of roughly 3.5 and 0.9, respectively. The physical reality is likely to be more complex than these simple linear relationships would imply. For example, Miller et al. (2005) noted an ambiguity between one- and two-line fits to the silicon-iron inferred dust abundances, suggesting the possibility of at least two regimes with different degrees of iron depletion versus silicon. Nevertheless, these linear trends are consistent with current ideas on silicate grain formation, whereby magnesium-bearing silicates tend to form roughly in concert with metallic iron in cool solar-composition stellar atmospheres and iron-bearing silicates require further processing to produce (Rietmeijer, Nuth, & Karner 1999; Whittet 2003). Using the simple linear relations derived from the dust ratio plots, accepted cosmic abundance standards can be compared for consistency with our data and known features of dust.

[5]The Miller et al. (2005) data has been refit using the Lodders (2003) solar photosphere for reference rather than the proto-Sun.

Sofia & Meyer (2001) recently re-evaluated the relevance of commonly-applied interstellar elemental abundance standards; specifically, these standards are abundances in B stars, young F and G stars, and the Sun. We make a similar comparison between them here, including protosolar abundances as a further alternative (Lodders 2003). Table 9 presents the features of each reference in terms of the total elemental abundances and the dust values inferred from the observed gas level means. The last row is the crux of the test, the amount of oxygen predicted to be found in dust, given the inferred magnesium, iron, and silicon amounts and a few simple assumptions regarding grain composition. First among these assumptions is that the linear relationships between magnesium or iron and silicon dust abundances produce reliable estimates of the amount of the former elements not involved in silicon compounds by way of the $N(\mathrm{Si})_{dust} = 0$ intercept. Assuming that MgO is the dominant form of magnesium dust not involving silicon and that $Fe_3O_4$ is the prevailing iron oxide form (Jones 1990), then each non-silicate Mg atom implies one O atom and every three Fe atoms yield 4 O atoms in dust. Next, assuming a 15:85 distribution ratio between pyroxene- and olivine-like magnesium and iron compounds of silicon (Kemper, Vriend, & Tielens 2004) specifies the amount of dust O for each atom of silicon depleted from the gas phase. Consequently, an approximate oxygen dust census can be derived from the mean magnesium, iron, and silicon abundances for high density sight lines.

The high-$\langle n_{\mathrm{H}} \rangle$ silicon dust abundances are identical with the means of values inferred from magnesium and iron dust abundances and the derived linear dust-to-dust relations for each cosmic standard. The predicted amount of oxygen in dust form, however, is not quite consistent with any of the standards. For B stars, silicate grains fulfil the oxygen allowance by themselves, not leaving any oxygen in dust for the magnesium and iron tied up in oxides. In contrast, the oxygen dust abundances that each of the solar references imply are too large to be accounted for by simple silicates and oxides built out of magnesium and iron. Nor is it likely that other abundant elements can take up the slack; the interstellar gas-phase carbon abundance, in terms of the C/H ratio, is essentially constant for diffuse and translucent sight lines (Sofia et al. 2004) whereas the corresponding oxygen value changes significantly. The young F and G star standard compiled by Sofia & Meyer (2001), however, comes quite close to matching the inferred oxygen dust abundance with the prediction based on magnesium, iron, and silicon. In fact, if FeO rather than $Fe_3O_4$ is adopted as the dominant reservoir of iron depletion outside of silicates, then the gap between the oxygen dust values is reduced by a factor of nearly two.[6] Consequently, the observed oxygen, magnesium, iron, and silicon abundances for high mean density sight lines appear to favor a cosmic standard close to that of young F and G stars with perhaps a slightly higher oxygen abundance. To inject a note of caution, however, it should be recalled that the standards that Sofia & Meyer (2001) derived from stellar atmosphere abundances have large uncertainties. Low-$\langle n_{\rm H} \rangle$ (warm ISM) depletion levels have not been included in the comparison of cosmic standards since the silicon dust abundances for each reference abundance level do not agree with those derived from magnesium and iron inferred dust abundances using the fitted linear relations. As suggested earlier, these simple correlations between magnesium, iron, and silicon dust quantities apparently do not extend to indefinitely small amounts of depletion.

## 5. Concluding Remarks

The elemental abundance measurements we have presented are based on high-quality data from GHRS, STIS, and *FUSE*; consequently, they describe density-dependent depletion more precisely than previous *Copernicus* and *IUE* datasets. Improvements in spectral resolution and sensitivity have made weak absorption features much more accessible, mitigating concerns about unresolved saturation. As a result, the trends in gas-phase elemental abundance as a function of mean hydrogen sight line density $\langle n_{\rm H} \rangle$ stand out starkly as transitions between two depletion plateaus associated with the warm neutral medium and cold clouds. Although the character of the transition changes with the degree to which each element is depleted, there is evidence for a common distinction between abundances for sight lines sparser and denser on average than about 1.5 cm$^{-3}$. Unfortunately, the quality of abundance measurements is not yet sufficiently improved to resolve the question around the existence of a possible third plateau in abundance at moderate values of $\langle n_{\rm H} \rangle$, with a second transition near 0.3 cm$^{-3}$. Nevertheless, the breadth of the data distribution for each element around the depletion function fit to it is sufficiently small to conclude that total elemental abundances are homogeneous to the limits of measurement uncertainty on length scales of hundreds of parsecs. Based on the scatter for undepleted krypton, a hard limit of 0.06 dex on intrinsic variability is evident, although the confluence of data from the eight elements we have studied suggest a more probable upper limit of 0.04 dex. Finally, the observed abundances and simple dust composition arguements tend to favor young F and G star atmospheric abundances as the cosmic reference standard implied by these limits on total abundance fluctuations.

[6]Although iron is found in a variety of other forms in meteorites, in the diffuse ISM metallic grains are likely to have been converted into oxides (Jones 1990) and sulphur depletion is minimal (e.g., Lehner, Wakker, & Savage 2004), precluding a large population of iron sulfides.

---

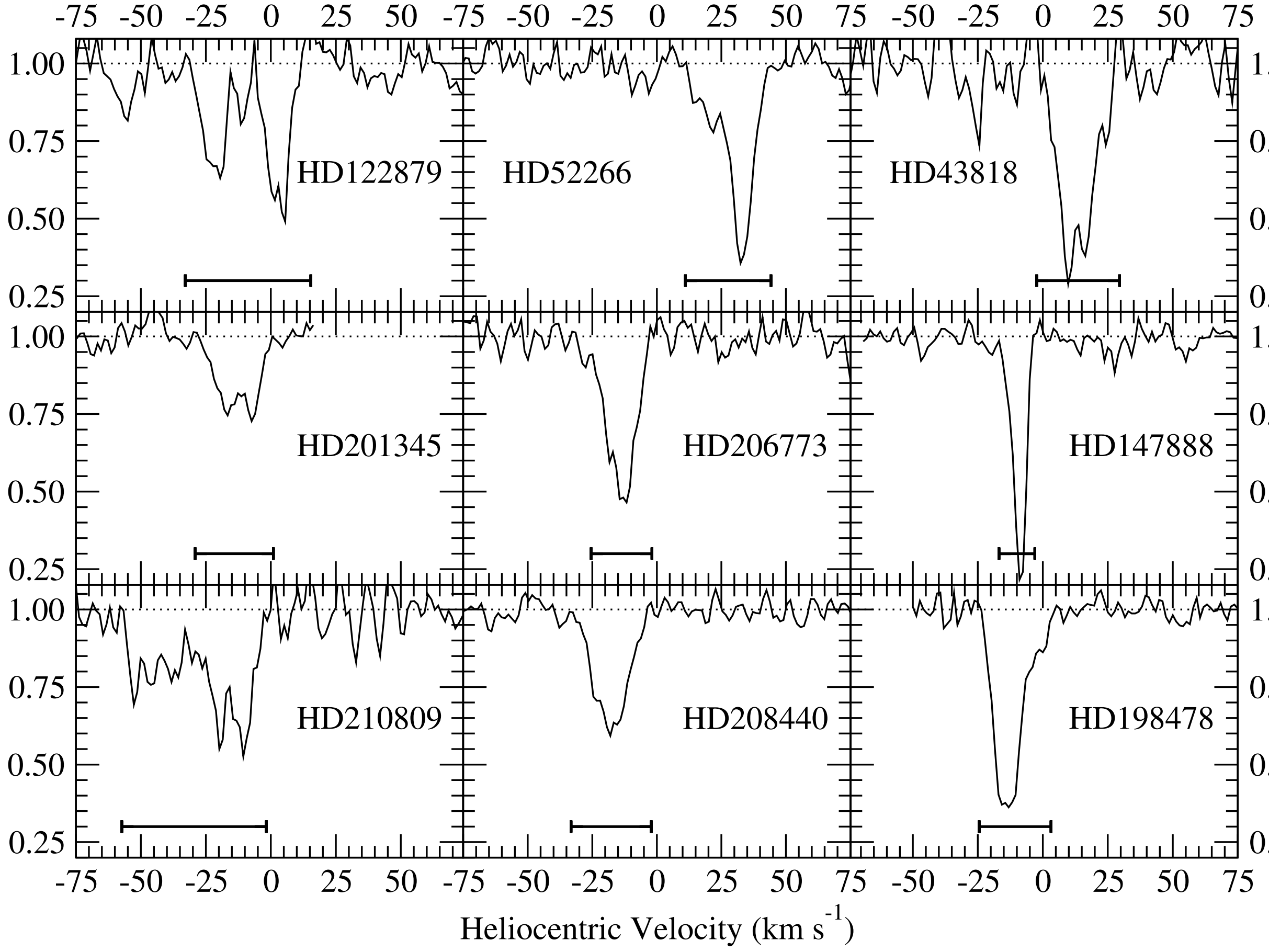

Fig. 1.— Characteristic absorption profiles for Ge II $\lambda$1237. A selection of normalized Ge II $\lambda$1237 absorption profiles are plotted above as a function of the heliocentric velocity. In each panel, the velocity range encompassed by the absorption features is identified by the line along the bottom axis. These widths were determined through consideration of the profiles for the dominant ionization states of krypton, oxygen, magnesium, phosphorus, manganese, nickel, copper, and germanium; other lines used in an advisory role included C I, S I, and Cl I transitions in the 1170 – 1372Å window. These panels show profiles characteristic of the diversity present in the current sample, which includes paths with as few as 1 or 2 absorption components, as for HD147888, and as many as 10 or more, as for HD210809.

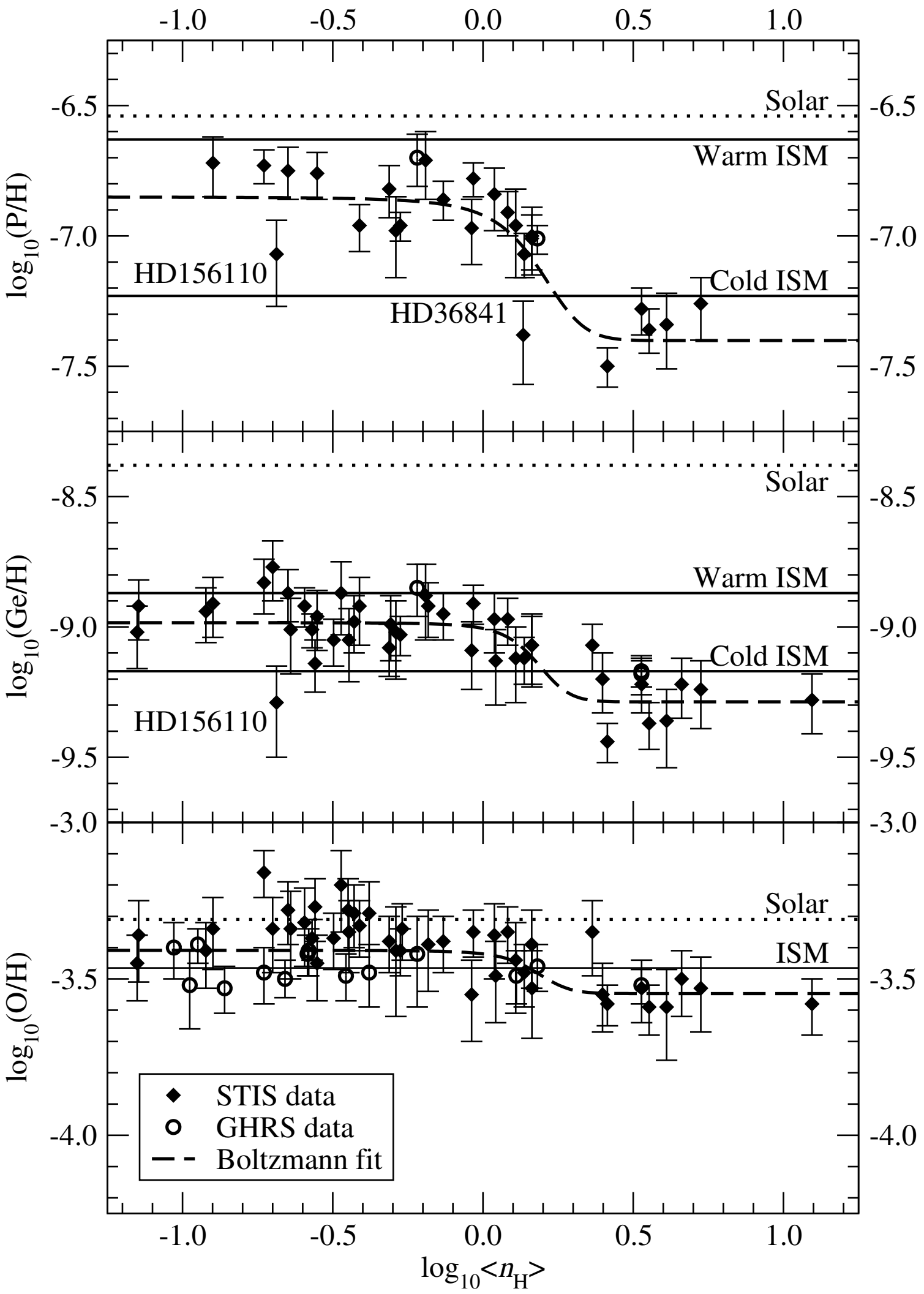

Fig. 2.— Density-dependent phosphorus, germanium, and oxygen abundances. Phosphorus and germanium gas-phase abundance ratios to hydrogen are plotted above as a function of $\langle n_{\rm H} \rangle$; oxygen abundances (Cartledge et al. 2004) are plotted in the bottom panel for comparison. The Boltzmann function we have adopted to describe the variation in elemental abundances is consistent with the distribution of each data set, allowing for the possibility that the HD1561110 sight line has a unique nature. The adopted solar abundance levels are taken from Lodders (2003) and the warm and cold neutral ISM gas-phase abundances are derived from Joseph (1993) in the case of phosphorus and Welty et al. (1999) for germanium, both adjusted for the change in solar references. The oxygen ISM reference level is based on a GHRS-only sample (Meyer, Jura, & Cardelli 1998). The phosphorus GHRS data are for $\xi$ Per (Cardelli et al. 1991) and 23 Ori (Welty et al. 1999); the germanium GHRS data represent 23 Ori, 1 Sco (Hobbs et al. 1993), and $\zeta$ Oph (Savage, Cardelli, & Sofia 1992). In each case that it was necessary, the GHRS measurements were updated to reflect the $f$-values used in this paper.

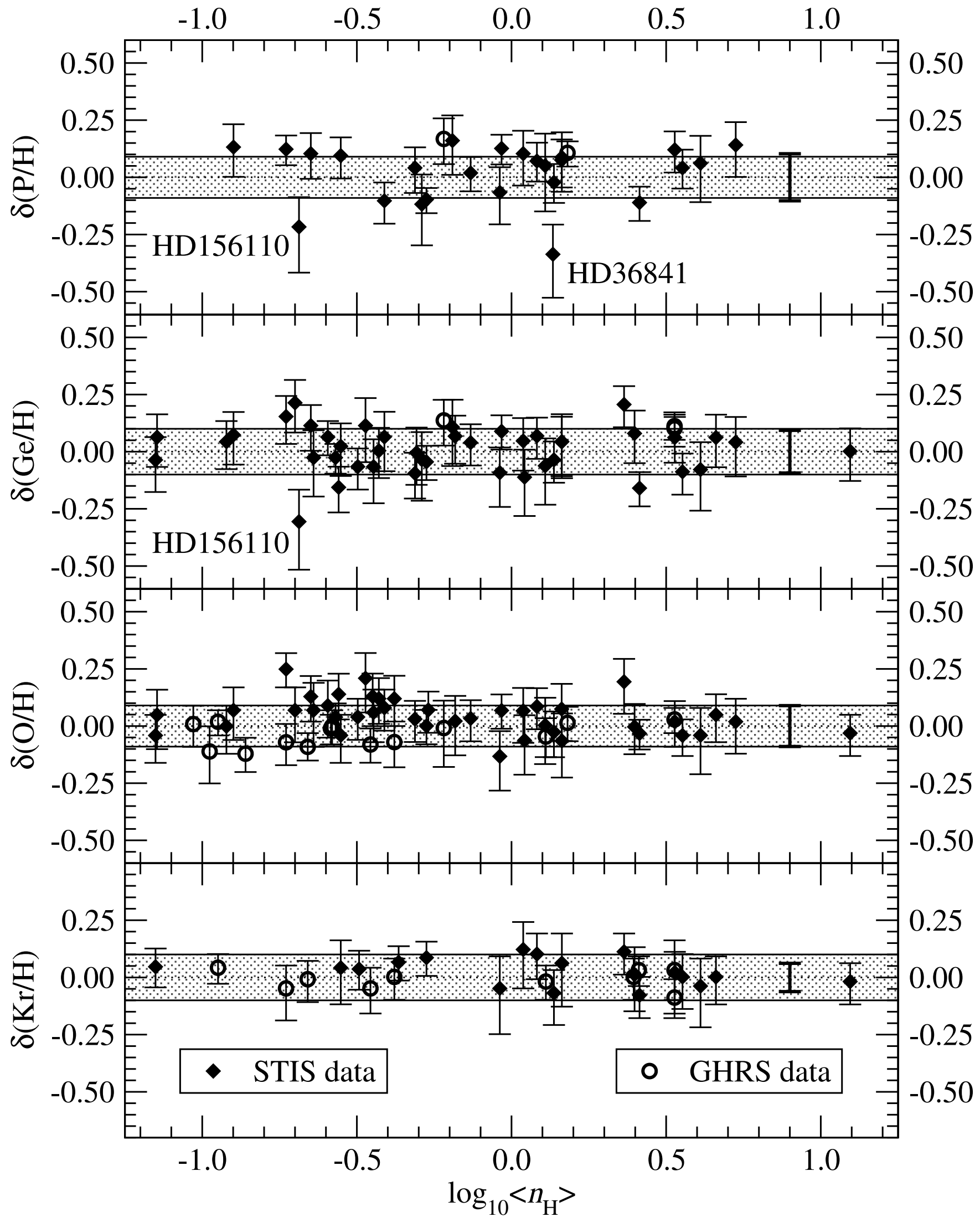

Fig. 3.— Scatter plot for phosphorus, germanium, oxygen, and krypton data with respect to Boltzmann functions (or a constant for krypton) fitted to each dataset. The shaded region in each panel indicates the bounds within which approximately 67% of the data should be found, assuming a gaussian distribution about the fitted function with breadth defined by a typical measurement uncertainty. The outlying phosphorus data points correspond to HD156110 (also a germanium outlier) and HD36841 (not detected in germanium). The bold error bars at $\log_{10}\langle n_{\rm H}\rangle = 0.9$ represent the measured width of each distribution; for phosphorus the leftward bars exclude HD156110 and HD36841. Also of note, the oxygen low-density data are influenced by the heliocentric distance effect (see § 3.3.1) and the sight line with enhanced germanium and oxygen abundances at $\log_{10}\langle n_{\rm H}\rangle = 0.36$ lies in the direction of HD37367. The oxygen data are taken from Cartledge et al. (2004) and the krypton results from Cartledge et al. (2006).

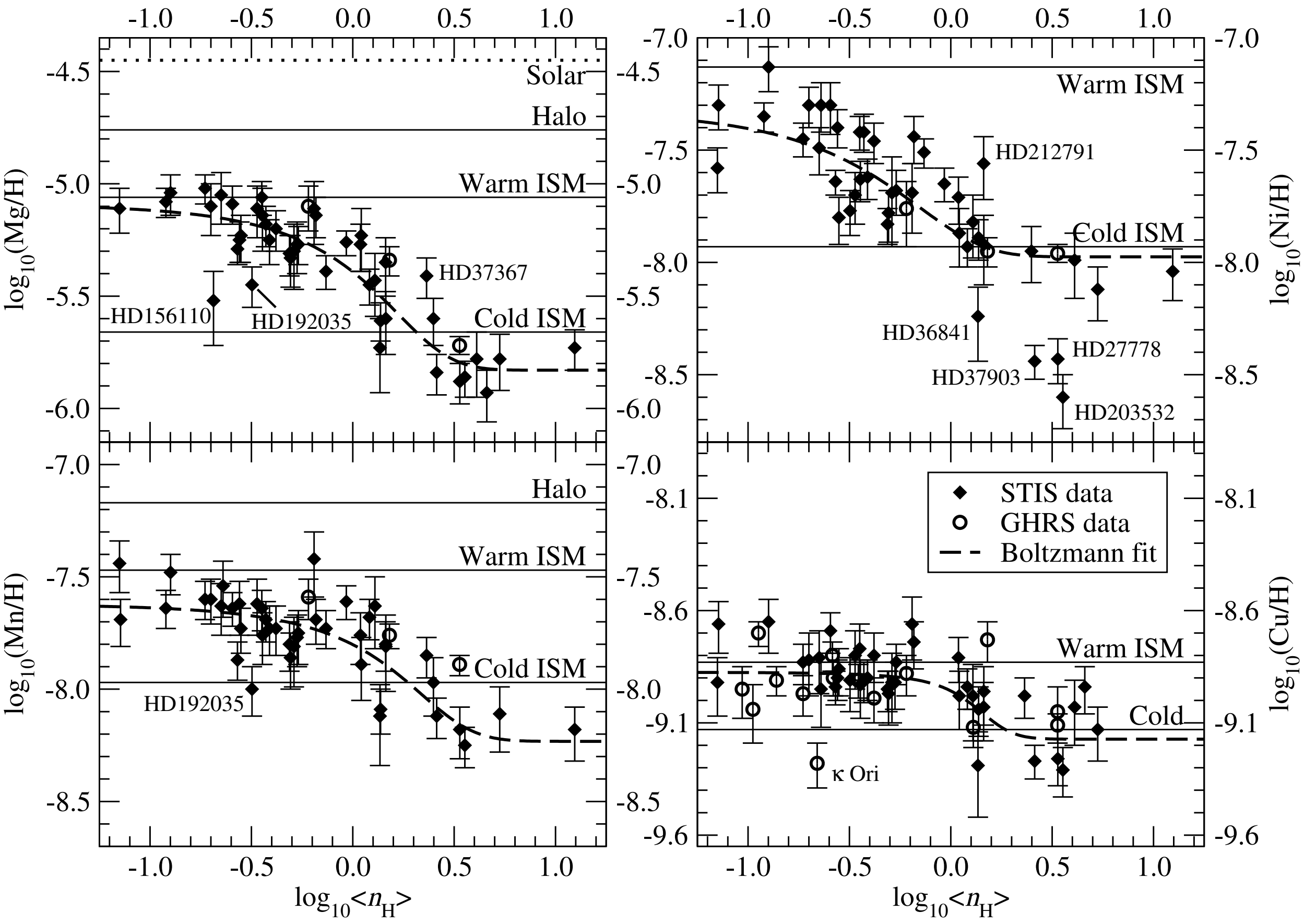

Fig. 4.— Density-dependent magnesium, manganese, copper, and nickel abundances. Gas-phase abundance ratios to hydrogen for heavily-depleted elements are plotted above as a function of $\langle n_{\rm H} \rangle$. Although the Boltzmann function description of abundance variation is generally consistent with each dataset, the scatter with respect to the fits is significantly increased relative to values measured in lightly-depleted elements. The broader distribution for magnesium, manganese, and nickel qualitatively suggests that an additional depletion plateau may be present, obscured by measurement and distance uncertainties. A possible lower branch to the nickel distribution is evident in the relatively low abundances of four sight lines which in other elements are consistent with the depletions of similarly-dense paths. The GHRS data plotted in this figure depict magnesium, manganese, nickel, and copper measurements for $\xi$ Per, 23 Ori, and $\zeta$ Oph, as well as unpublished interstellar copper data from $\gamma$ Cas, $\zeta$ Per, $\epsilon$ Per, $\delta$ OriA, $\lambda$ Ori, $\iota$ Ori, $\epsilon$ Ori, $\kappa$ Ori, 15 Mon, $\tau$ CMa, and $\gamma$ Ara (measured by D. M.). The plotted abundances references are drawn from Lodders (2003) (solar) and Welty et al. (1999) (all others).

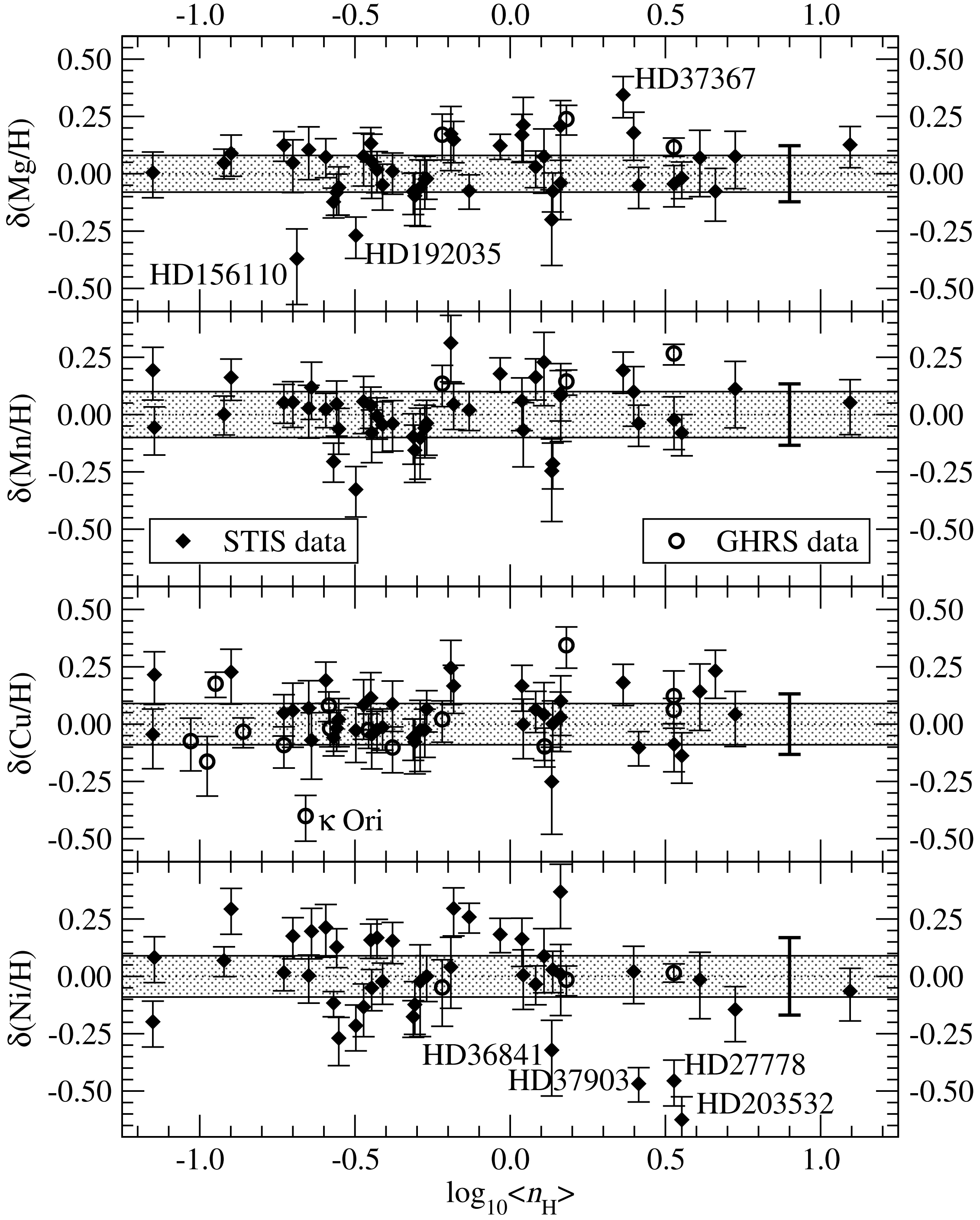

Fig. 5.— Scatter plot for magnesium, manganese, copper, and nickel data with respect to Boltzmann functions fitted to each dataset. As in Figure 3, the shaded region and bold error bar in each panel identify the relative size of abundance ratio uncertainty and measured scatter. Noted outliers common to more than one element include HD36841 (at $\log_{10}\langle n_{\rm H}\rangle = 0.13$), HD37367 (at 0.36), and HD192035 (at $-0.50$). Other notable excursions include $\kappa$ Ori in copper and HD27778, HD37903, and HD203532 in nickel.

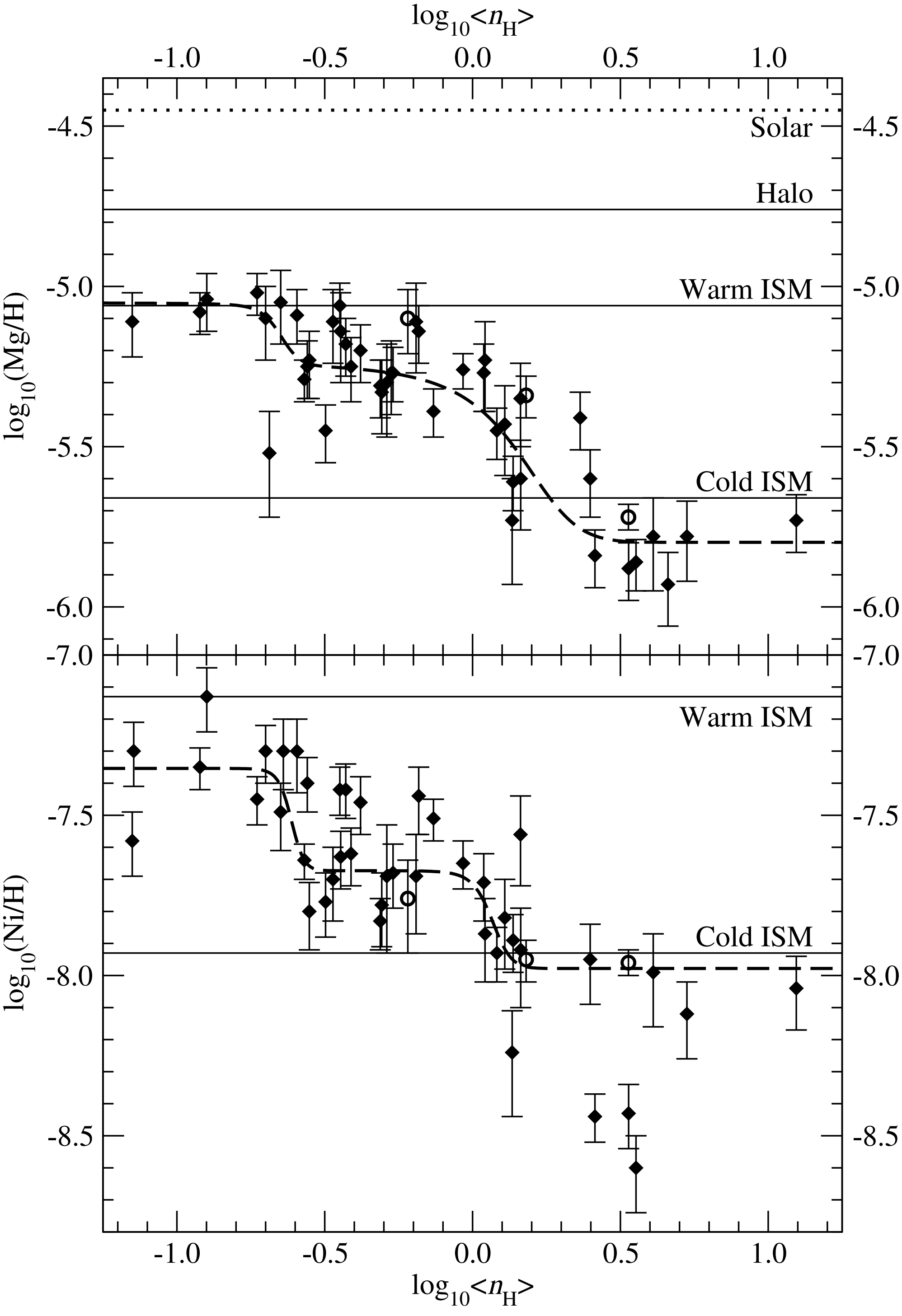

Fig. 6.— Density-dependent magnesium and nickel abundances fit by Boltzmann functions with three plateaus. The gas-phase abundances presented in Figure 4 for magnesium and nickel are plotted above and fit by a function allowing for three abundance plateaus and two transitions between them (formally defined in footnote $a$ to Table 6). These fits aesthetically improve the quality of the fit with respect to the two-plateau hypothesis; however, $\chi^2_\nu$ and the scatter are not reduced.

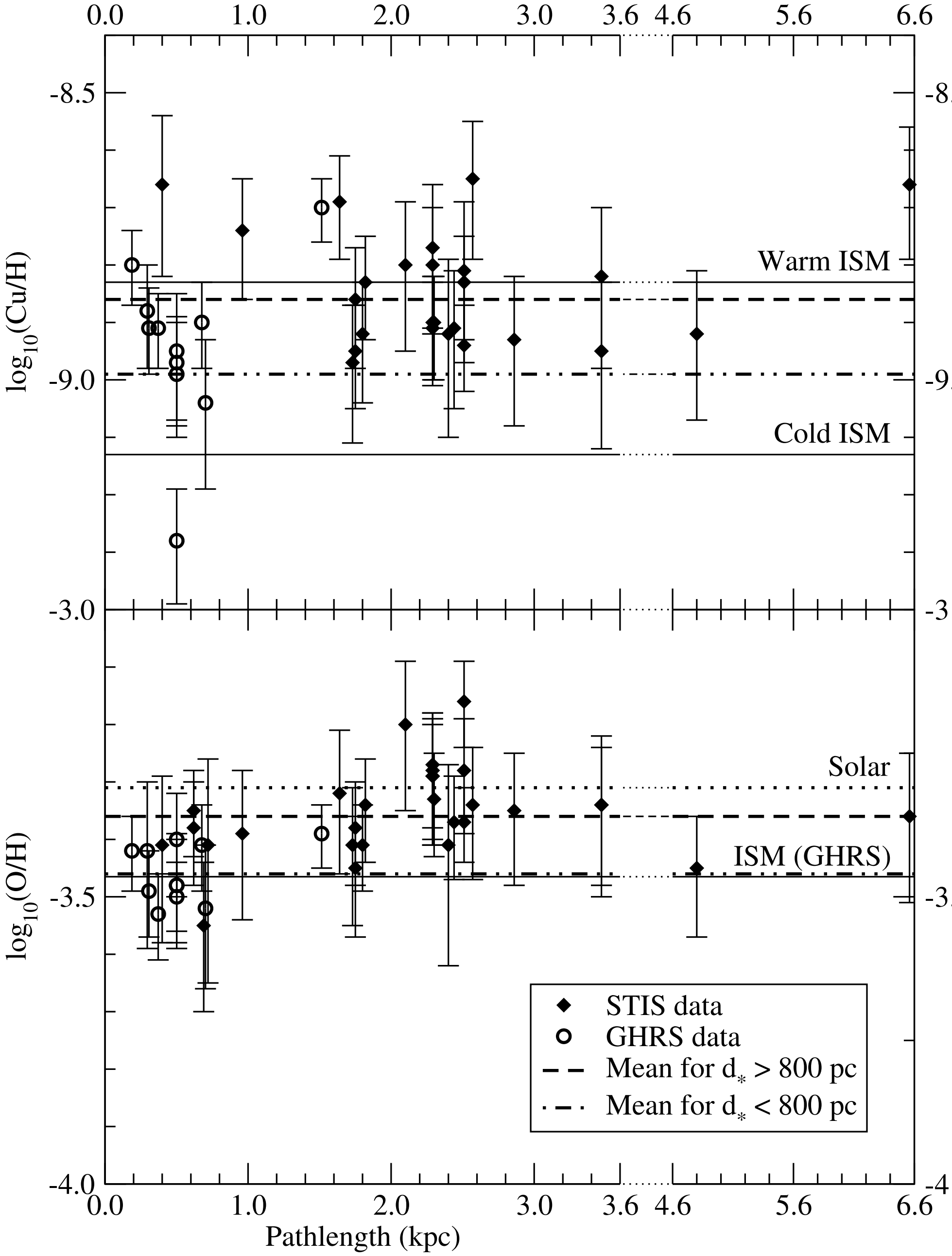

Fig. 7.— Copper and oxygen gas-phase abundances for low-$\langle n_{\rm H}\rangle$ sight lines plotted as a function of pathlength. Copper and oxygen provide the best evidence for reduced elemental abundances for low density ($\langle n_{\rm H}\rangle < 1.0$ cm$^{-3}$ or $\log_{10}\langle n_{\rm H}\rangle < 0.0$) sight lines shorter than 800 pc relative to those typical of longer paths. The gaps between the short- and long-path means, $0.129\pm0.033$ dex and $0.099\pm0.028$ dex and for copper and oxygen respectively, match within error and are thus consistent with the possibility that they are explained by a recent low-metallicity gas infall onto the Galactic disk. The current database does not include sufficient short path measurements to establish the effect for other elements.

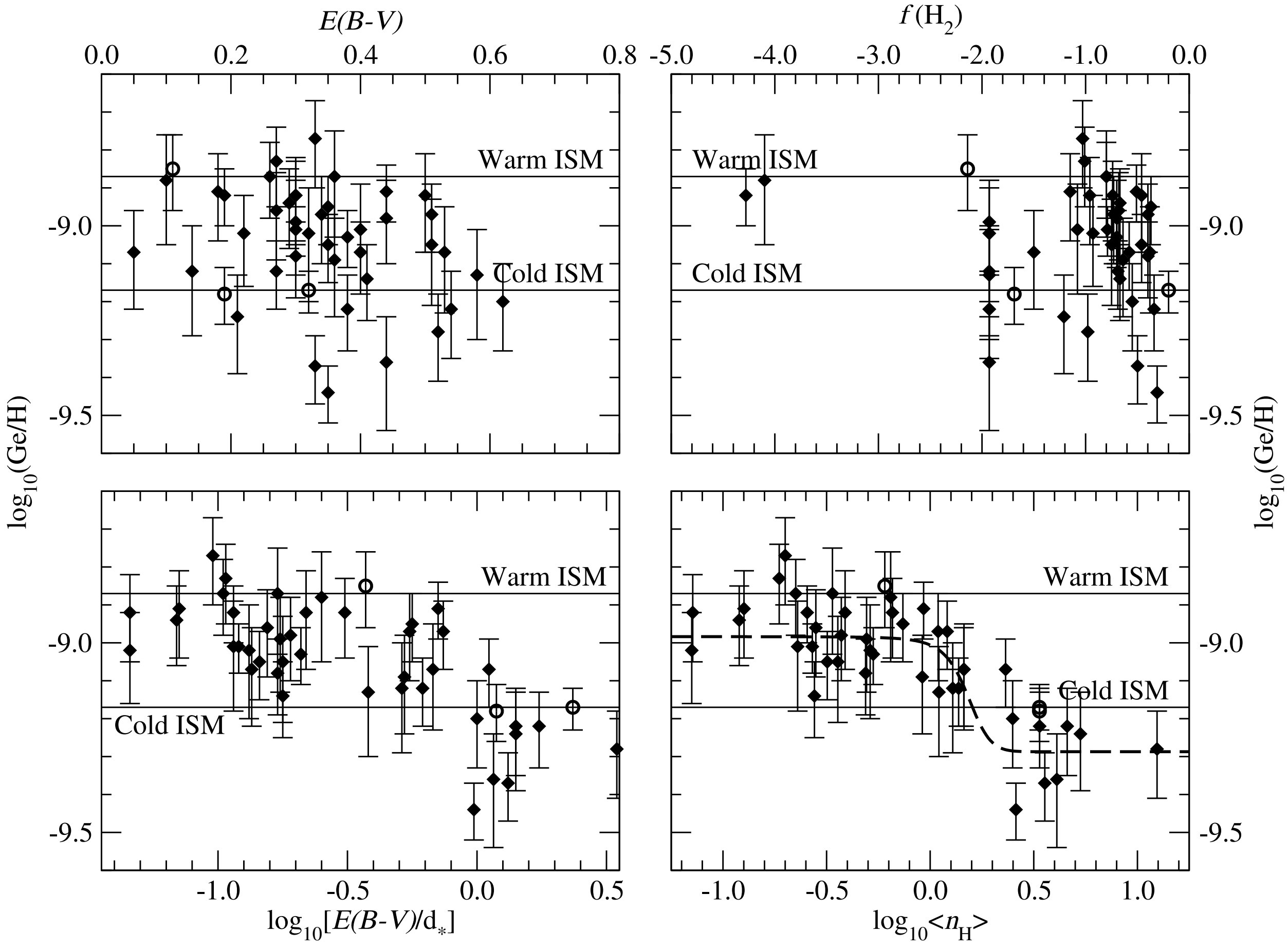

Fig. 8.— Germanium gas-phase abundances expressed relative to $E(B-V)$, $E(B-V)/d_*$, $f(\mathrm{H}_2)$, and $\langle n_\mathrm{H} \rangle$. Germanium has been selected to test the gas-phase abundance variations for lightly-depleted elements with sight line properties other than mean hydrogen sight line density. Among these, only $E(B-V)/d_*$ allows a capability for determining distinct depletion levels comparable to that available using $\langle n_\mathrm{H} \rangle$. With the density parameter, however, scatter relative to the fitting function is smaller, indicating that it is the better discriminating factor. As with $E(B-V)$ and $f(\mathrm{H}_2)$, no helpful trends are evident when abundances are plotted as functions of sky position or $R_V$. HD156110 has been omitted from this figure.

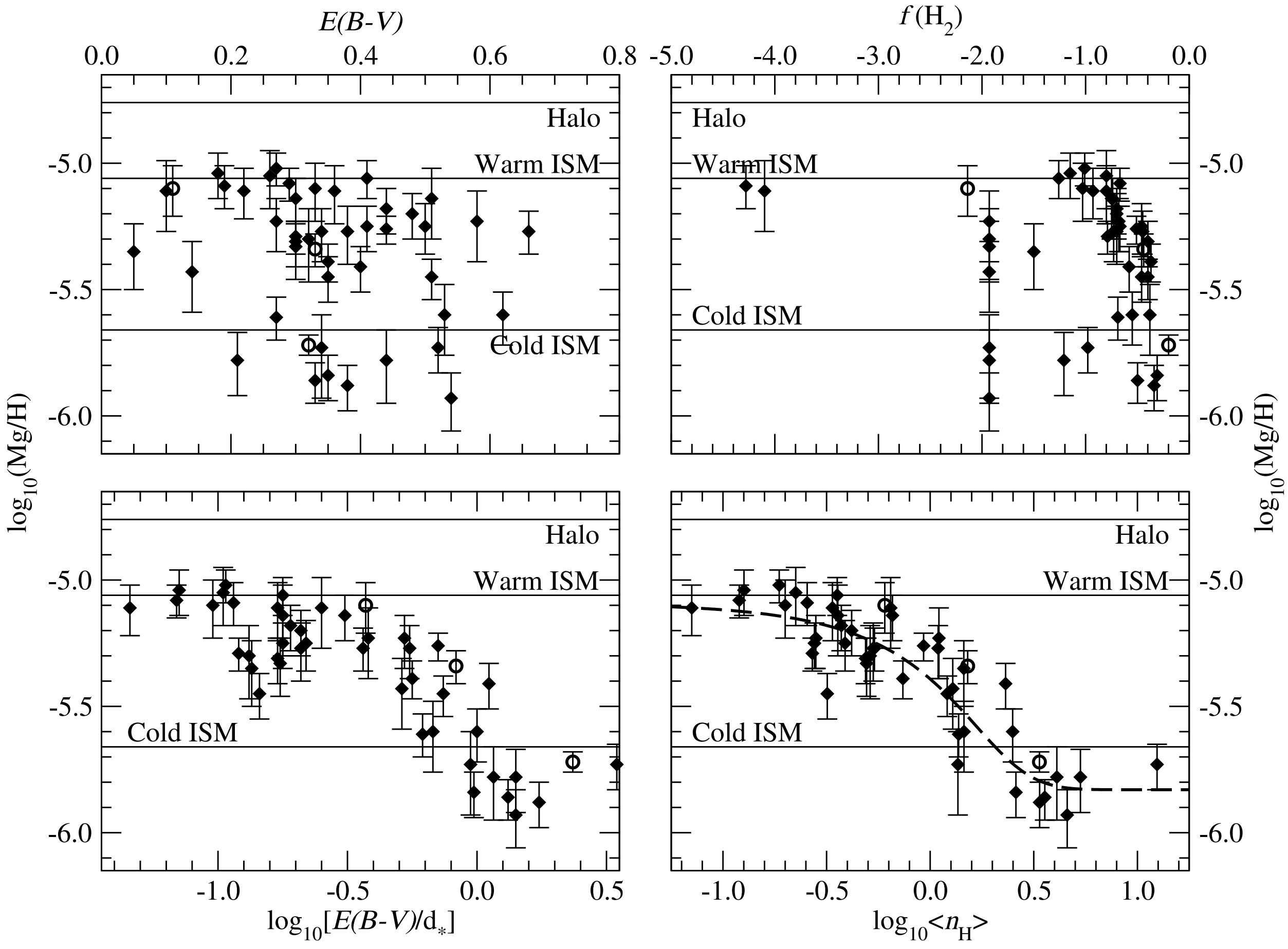

Fig. 9.— Magnesium gas-phase abundances expressed relative to $E(B-V)$, $E(B-V)/d_*$, $f(H_2)$, and $\langle n_H \rangle$. Magnesium has been selected to test the gas-phase abundance variations for more heavily depleted elements with sight line properties other than mean hydrogen sight line density. As with germanium, only $E(B-V)/d_*$ approaches the effectiveness of $\langle n_H \rangle$. HD156110 has been omitted from this figure.

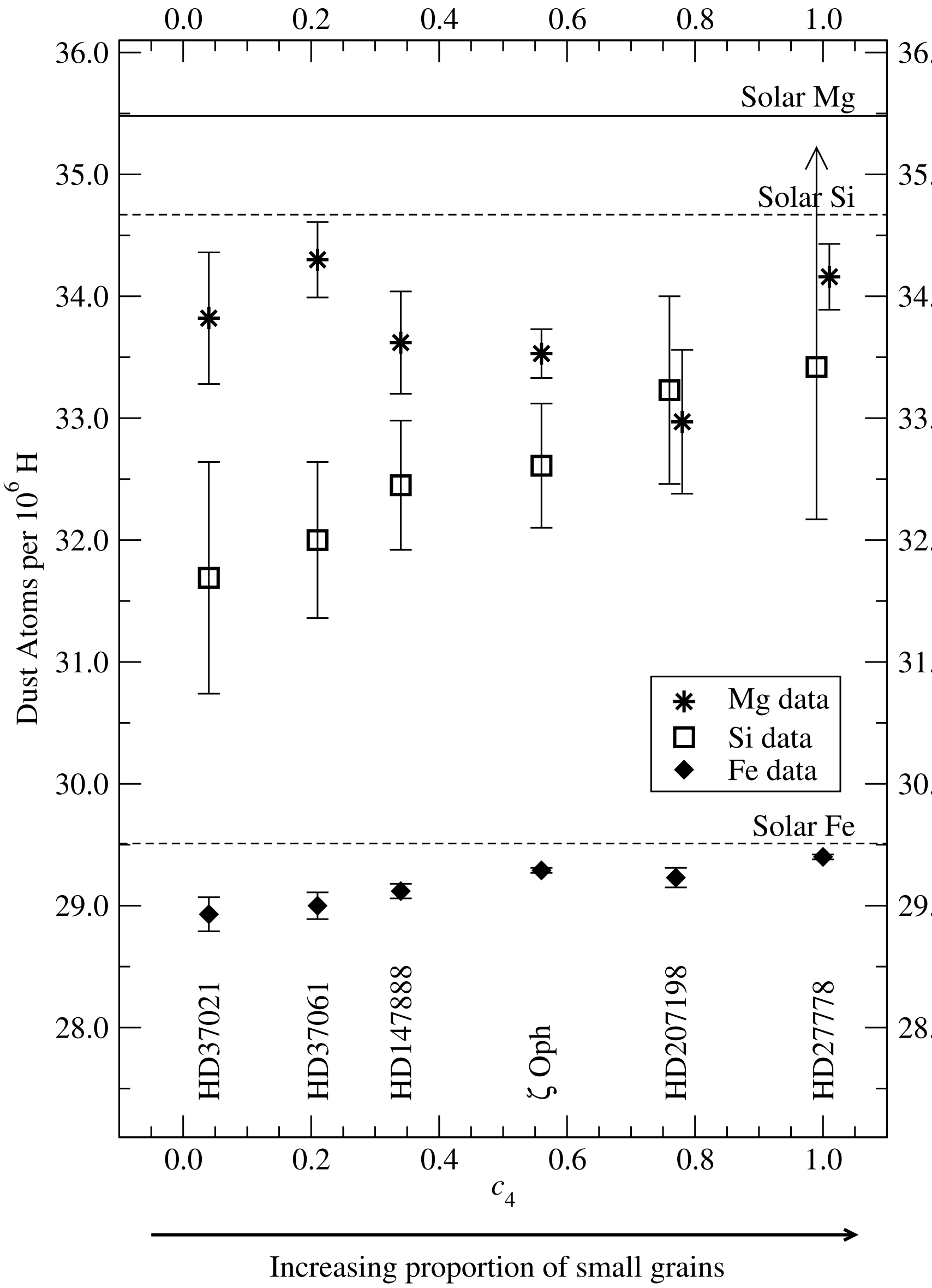

Fig. 10.— Magnesium, silicon, and iron dust abundance variations with extinction curve parameter $c_4$ (Fitzpatrick & Massa 1990; Valencic et al. 2004). This plot combines Miller et al. (2005) iron and silicon results with magnesium data for translucent sight lines in the current sample; the points representing Si and Mg toward HD27778 and HD207198 are offset for clarity. The trend for both iron and silicon is an increasing dust abundance with $c_4$, a property that rises as small grains make up a larger fraction of the dust population. In contrast, magnesium abundances are relatively constant, or appear to decrease if HD27778 is neglected. HD27778 is unique among these paths for its very low nickel and carbon gas-phase abundances (§ 3.2; Sofia et al. 2004). Lodders (2003) photospheric abundances were adopted as the cosmic standard; Savage, Cardelli, & Sofia (1992) provided the $\zeta$Oph data, which were adjusted to reflect currently-accepted $f$-values.

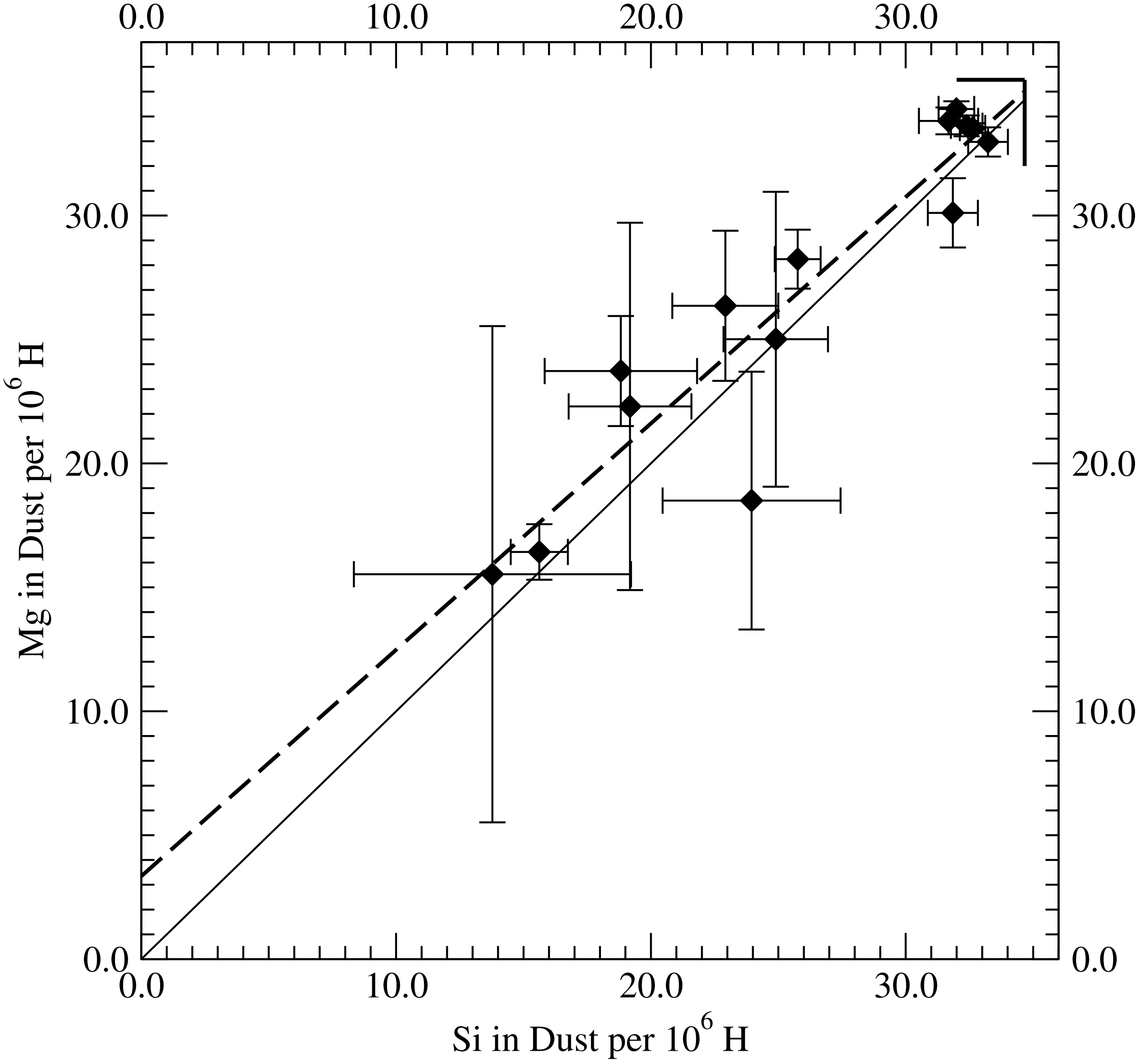

Fig. 11.— A comparison of magnesium and silicon dust abundances. For sight lines probing a variety of interstellar environments and their corresponding depletion levels, the inferred magnesium and silicon dust abundances are consistent with a constant ratio of 0.91 after an initial depletion of four magnesium atoms per million H for no silicon in dust. The Lodders (2003) photospheric references are indicated by the edges in the upper right corner. The dashed line is the least-squares fit to the data; the solid line indicates a one-to-one dust ratio.

Table 1. Summary of Observations

| Sight Line | *STIS* Data Set | Date | Time (s) | *FUSE* Data Set | Date | Time (s) |
|---|---|---|---|---|---|---|
| HD1383 | O5C07C010 | 1999 Nov 08 | 1440 | B0710101000 | 2000 Sep 30 | 23332 |
| HD12323 | O63505010 | 2001 Apr 02 | 1440 | P1020202000 | 1999 Nov 25 | 3865 |
| HD13268 | O63506010 | 2001 Feb 01 | 1440 | P1020304000 | 1999 Nov 24 | 4438 |
| HD14434 | O63508010 | 2001 Apr 06 | 1440 | P1020504000 | 1999 Nov 24 | 4441 |
| HD27778 | O59S01010 | 2000 Mar 29 | 1657 | P1160301000 | 2000 Oct 27 | 9729 |
| | O59S01020 | 2000 Mar 30 | 2656 | — | — | — |
| HD36841 | O63516010 | 2000 Nov 24 | 1440 | — | — | — |
| HD37021 | O59S02010 | 2000 Jan 23 | 1643 | — | — | — |
| HD37061 | O59S03010 | 2000 Apr 02 | 1087 | — | — | — |
| HD37367 | O5C013010 | 2000 Mar 31 | 360 | — | — | — |
| HD37903 | O59S04010 | 2000 Feb 21 | 1643 | P1160601000 | 2001 Oct 18 | 4040 |
| HD43818 | O5C07I010 | 2001 Apr 10 | 1440 | — | — | — |
| HD52266 | O5C027010 | 2000 Mar 11 | 360 | — | — | — |
| HD63005 | O63531010 | 2001 May 07 | 1440 | P1022101000 | 2000 Apr 05 | 5311 |
| HD71634 | O5C090010 | 2000 May 22 | 720 | Z9012201000 | 2002 Mar 08 | 504 |
| HD72754 | O5C03E010 | 1999 Oct 14 | 720 | B0710302000 | 2002 May 04 | 5950 |
| HD75309 | O5C05B010 | 2000 Mar 28 | 720 | P1022701000 | 2000 Jan 26 | 4684 |
| HD79186 | O5C092010 | 2000 May 25 | 360 | P2310301000 | 2002 May 04 | 5393 |
| HD91824 | O5C095010 | 2000 Mar 23 | 360 | A1180802000 | 2000 Feb 06 | 4649 |
| HD91983 | O5C08N010 | 2000 May 18 | 1440 | B0710402000 | 2001 May 27 | 4308 |
| HD111934 | O5C03N010 | 2000 Mar 27 | 720 | — | — | — |
| HD116852 | O5C01C010 | 2000 Jun 25 | 360 | P1013801000 | 2000 May 27 | 7212 |
| HD122879 | O5C037010 | 2000 May 08 | 360 | B0710501000 | 2002 Mar 03 | 1532 |
| HD147888 | O59S05010 | 2000 Aug 17 | 1656 | P1161501000 | 2003 Aug 21 | 12205 |
| HD148594 | O5C04A010 | 2000 May 21 | 720 | P2310101000 | 2001 Aug 19 | 4016 |
| HD152590 | O5C08P010 | 2000 Mar 03 | 1440 | B0710602000 | 2001 Aug 12 | 2773 |
| | — | — | — | B0710601000 | 2001 Jul 08 | 1389 |
| HD156110 | O5C01K010 | 1999 Nov 12 | 360 | B0710701000 | 2001 May 08 | 503 |
| HD157857 | O5C04D010 | 2000 Jun 03 | 720 | P1027501000 | 2000 Sep 02 | 4022 |

Table 1—Continued

| Sight Line | *STIS* Data Set | Date | Time (s) | *FUSE* Data Set | Date | Time (s) |
|---|---|---|---|---|---|---|
| HD165955 | O63599010 | 2001 May 11 | 1440 | P1027901000 | 2000 Aug 17 | 4140 |
| HD175360 | O5C047010 | 2000 Apr 26 | 360 | — | — | — |
| HD185418 | O5C01Q010 | 1999 Jul 17 | 720 | P1162301000 | 2000 Aug 10 | 4416 |
| HD190918 | O6359J010 | 2000 Oct 12 | 720 | P1028501000 | 2000 Jul 18 | 5510 |
| HD192035 | O6359K010 | 2001 Feb 21 | 1440 | P1028603000 | 2000 Jun 21 | 6624 |
| | — | — | — | P1028602000 | 2000 Jun 19 | 6148 |
| | — | — | — | P1028601000 | 2000 Jun 17 | 4942 |
| HD192639 | O5C08T010 | 2000 Mar 01 | 1440 | P1162401000 | 2000 Jun 12 | 4834 |
| HD198478 | O5C06J010 | 2000 Apr 04 | 720 | P2350201000 | 2001 Jun 06 | 6628 |
| HD198781 | O5C049010 | 1999 Sep 06 | 360 | P2310201000 | 2001 Jul 21 | 844 |
| HD201345 | O5C050010 | 1999 Nov 12 | 360 | P1223001000 | 2000 Jun 13 | 5104 |
| HD203532 | O5C01S010 | 2000 Jun 29 | 720 | B0710801000 | 2001 Aug 23 | 4748 |
| HD206773 | O5C04T010 | 1999 Aug 12 | 360 | B0710901000 | 2001 Jul 19 | 4399 |
| HD207198 | O59S06010 | 2000 Oct 30 | 1856 | P1162801000 | 2000 Jul 23 | 13180 |
| | O59S06020 | 2000 Oct 30 | 2855 | — | — | — |
| HD208440 | O5C06M010 | 1999 Nov 05 | 720 | B0300401000 | 2001 Aug 03 | 9777 |
| HD210809 | O5C01V010 | 1999 Oct 30 | 720 | P1223103000 | 2000 Aug 08 | 10065 |
| | O6359T010 | 2001 Apr 24 | 720 | P1223102000 | 2000 Aug 07 | 7097 |
| | — | — | — | P1223101000 | 2000 Aug 05 | 5508 |
| HD212791 | O5C04Q010 | 2000 Apr 25 | 1440 | Z9017701000 | 2002 Jun 30 | 2261 |
| HD220057 | O5C01X010 | 2000 Mar 24 | 360 | Z9017801000 | 2002 Aug 30 | 3037 |
| HD232522 | O5C08J010 | 1999 Oct 09 | 1440 | P1220101000 | 1999 Nov 28 | 16800 |
| HD308813 | O63559010 | 2001 Feb 18 | 1440 | P1221903000 | 2000 Mar 27 | 5667 |
| | — | — | — | P1221902000 | 2000 Mar 25 | 4340 |
| | — | — | — | P1221901000 | 2000 Mar 23 | 4257 |
| BD +53 2820 | O6359Q010 | 2001 Feb 05 | 1440 | M1141301000 | 2000 Aug 06 | 2529 |
| | — | — | — | P1223203000 | 2000 Aug 08 | 5147 |
| | — | — | — | P1223202000 | 2000 Aug 07 | 7119 |
| | — | — | — | P1223201000 | 2000 Aug 06 | 5815 |

Table 1—Continued

| Sight Line | *STIS* Data Set | Date | Time (s) | *FUSE* Data Set | Date | Time (s) |
|---|---|---|---|---|---|---|
| CPD −69 1743 | O63566010 | 2001 Feb 20 | 1440 | P1013701000 | 2000 Apr 06 | 3567 |

Table 2. Absorption Line Properties

| Ion | $\lambda_{\rm vac}$ (Å) | $f$-value |
|---|---|---|
| | Key Lines | |
| Cu II | 1358.7730 | $3.803 \times 10^{-1}$ |
| Ge II | 1237.0591 | $1.232 \times 10^{0}$ |
| Kr I | 1235.8380 | $2.042 \times 10^{-1}$ |
| Mg II | 1239.9253 | $6.318 \times 10^{-4}$ |
| | 1240.3947 | $3.560 \times 10^{-4}$ |
| Mn II | 1197.184 | $1.566 \times 10^{-1}$ |
| Ni II | 1317.217 | $7.750 \times 10^{-2}$ |
| O I | 1355.5977 | $1.160 \times 10^{-6}$ |
| P II | 1301.840 | $1.725 \times 10^{-2}$ |
| | Other Lines | |
| C I | 1276.4825 | $4.490 \times 10^{-3}$ |
| | 1277.2454 | $9.350 \times 10^{-2}$ |
| | 1280.1353 | $2.290 \times 10^{-2}$ |
| | 1328.8333 | $6.300 \times 10^{-2}$ |
| C I* | 1276.7498 | $2.520 \times 10^{-3}$ |
| | 1277.2823 | $7.060 \times 10^{-2}$ |
| | 1277.5130 | $2.230 \times 10^{-2}$ |
| | 1279.8904 | $1.260 \times 10^{-2}$ |
| | 1280.4042 | $4.260 \times 10^{-3}$ |
| | 1280.5970 | $6.740 \times 10^{-3}$ |
| C I** | 1277.5496 | $7.920 \times 10^{-2}$ |
| | 1277.7229 | $1.550 \times 10^{-2}$ |
| | 1277.9538 | $8.180 \times 10^{-4}$ |
| | 1280.3328 | $1.420 \times 10^{-2}$ |
| | 1280.8470 | $4.920 \times 10^{-3}$ |
| Cl I | 1347.2396 | $1.530 \times 10^{-1}$ |
| Ga II | 1414.402 | $1.799 \times 10^{0}$ |
| Ni II | 1345.878 | $7.640 \times 10^{-3}$ |

Table 2—Continued

| Ion | $\lambda_{vac}$ (Å) | $f$-value |
|---|---|---|
| | 1370.132 | $7.640 \times 10^{-2}$ |
| S I | 1247.1602 | $4.428 \times 10^{-3}$ |
| | 1270.7686 | $1.104 \times 10^{-4}$ |
| | 1270.7804 | $8.611 \times 10^{-3}$ |
| | 1270.7874 | $1.622 \times 10^{-3}$ |
| | 1295.6531 | $8.700 \times 10^{-2}$ |
| | 1296.1739 | $2.200 \times 10^{-2}$ |
| | 1303.4300 | $4.508 \times 10^{-3}$ |
| | 1316.5425 | $2.871 \times 10^{-2}$ |
| | 1316.6150 | $5.458 \times 10^{-3}$ |
| | 1316.6219 | $3.690 \times 10^{-4}$ |

Note. — The adopted $f$-values are from Morton (1991), Biémont, Morton, & Quinet (1998), and Welty et al. (1999), and are used for both apparent-optical-depth and profile-fitting elemental-abundance measurements.

Table 3. Elemental Measurements

| Star | $W_\lambda$ | $\log_{10}(N)_A$ | $\log_{10}(N)_P$ | $\log_{10}(N)$[a] |
|---|---|---|---|---|
| | | *Magnesium: Mg II λ1240 Doublet*[b] | | |
| HD1383 | 122.9 (4.2) | 16.39 (0.03) | 16.36 (0.08) | 16.36 (0.09) |
| HD12323 | 65.7 (2.2) | 16.03 (0.02) | 16.04 (0.05) | 16.04 (0.05) |
| HD13268 | 98.4 (4.9) | 16.24 (0.03) | 16.24 (0.04) | 16.24 (0.05) |
| HD14434 | 120.7 (6.8) | 16.26 (0.03) | 16.27 (0.04) | 16.27 (0.05) |
| HD27778 | 18.1 (0.7) | 15.49 (0.02) | 15.48 (0.01) | 15.48 (0.02) |
| HD36841 | 20.9 (2.0) | 15.45 (0.07) | 15.45 (0.02) | 15.45 (0.07) |
| HD37021 | 33.9 (0.9) | 15.85 (0.02) | 15.90 (0.02) | 15.90 (0.03) |
| HD37061 | 27.6 (0.5) | 15.78 (0.01) | 15.80 (0.05) | 15.80 (0.05) |
| HD37367 | 46.4 (1.0) | 15.99 (0.02) | 16.00 (0.03) | 16.00 (0.04) |
| HD37903 | 22.2 (0.7) | 15.65 (0.02) | 15.62 (0.06) | 15.62 (0.06) |
| HD43818 | 97.7 (2.5) | 16.46 (0.02) | 16.48 (0.03) | 16.48 (0.04) |
| HD52266 | 62.0 (1.7) | 16.08 (0.02) | 16.09 (0.01) | 16.09 (0.02) |
| HD63005 | 53.5 (1.5) | 16.00 (0.02) | 16.03 (0.02) | 16.03 (0.03) |
| HD71634 | 29.1 (0.9) | 15.78 (0.04) | 15.79 (0.02) | 15.79 (0.04) |
| HD75309 | 50.8 (1.8) | 15.94 (0.02) | 15.95 (0.05) | 15.95 (0.05) |
| HD79186 | 55.8 (1.4) | 16.09 (0.02) | 16.11 (0.03) | 16.11 (0.04) |
| HD91824 | 71.9 (2.1) | 16.12 (0.02) | 16.14 (0.02) | 16.14 (0.03) |
| HD91983 | 74.9 (2.1) | 16.19 (0.02) | 16.19 (0.06) | 16.19 (0.06) |
| HD111934 | 85.0 (2.3) | 16.26 (0.02) | 16.28 (0.01) | 16.28 (0.02) |
| HD116852 | 55.1 (3.7) | 15.91 (0.04) | 15.91 (0.02) | 15.91 (0.04) |
| HD122879 | 90.2 (1.9) | 16.22 (0.02) | 16.23 (0.01) | 16.23 (0.02) |
| HD147888 | 23.3 (0.5) | 15.84 (0.03) | 16.03 (0.03) | 16.03 (0.04) |
| HD148594 | 20.8 (0.9) | 15.65 (0.03) | 15.61 (0.02) | 15.61 (0.04) |
| HD152590 | 64.9 (1.7) | 16.19 (0.02) | 16.20 (0.09) | 16.20 (0.09) |
| HD156110 | 10.9 (0.8) | 15.15 (0.03) | 15.14 (0.02) | 15.14 (0.04) |
| HD157857 | 67.8 (3.1) | 16.17 (0.03) | 16.19 (0.05) | 16.19 (0.06) |
| HD165955 | 61.7 (2.4) | 16.04 (0.02) | 16.02 (0.05) | 16.02 (0.05) |
| HD175360 | 24.9 (1.0) | 15.60 (0.02) | 15.60 (0.05) | 15.60 (0.05) |
| HD185418 | 40.6 (1.2) | 15.97 (0.02) | 15.96 (0.01) | 15.96 (0.02) |

Table 3—Continued

| Star | $W_\lambda$ | $\log_{10}(N)_A$ | $\log_{10}(N)_P$ | $\log_{10}(N)$[a] |
|---|---|---|---|---|
| HD190918 | 123.1 (3.1) | 16.34 (0.03) | 16.34 (0.02) | 16.34 (0.04) |
| HD192035 | 50.4 (2.3) | 15.93 (0.03) | 15.93 (0.03) | 15.93 (0.04) |
| HD192639 | 62.2 (2.2) | 16.19 (0.02) | 16.21 (0.02) | 16.21 (0.03) |
| HD198478 | 45.9 (1.5) | 15.94 (0.02) | 15.95 (0.03) | 15.95 (0.04) |
| HD198781 | 40.7 (1.7) | 15.75 (0.03) | 15.76 (0.03) | 15.76 (0.04) |
| HD201345 | 55.9 (1.6) | 15.98 (0.02) | 15.96 (0.01) | 15.96 (0.03) |
| HD203532 | 21.5 (0.7) | 15.59 (0.02) | 15.58 (0.01) | 15.58 (0.02) |
| HD206773 | 53.0 (1.6) | 16.01 (0.02) | 15.99 (0.01) | 15.99 (0.02) |
| HD207198 | 54.5 (1.3) | 16.08 (0.01) | 16.08 (0.02) | 16.08 (0.02) |
| HD208440 | 59.9 (1.8) | 16.06 (0.03) | 16.05 (0.04) | 16.05 (0.05) |
| HD210809 | 95.6 (2.9) | 16.27 (0.02) | 16.23 (0.06) | 16.23 (0.06) |
| HD212791 | 49.5 (1.3) | 15.87 (0.02) | 15.87 (0.07) | 15.87 (0.07) |
| HD220057 | 28.2 (1.0) | 15.66 (0.02) | 15.66 (0.02) | 15.66 (0.03) |
| HD232522 | 65.0 (2.1) | 16.12 (0.02) | 16.11 (0.02) | 16.11 (0.03) |
| HD308813 | 86.8 (3.1) | 16.14 (0.02) | 16.15 (0.02) | 16.15 (0.03) |
| | | *Phosphorus: P II $\lambda$1301* | | |
| HD27778 | 22.3 (1.0) | 14.06 (0.02) | 14.08 (0.02) | 14.08 (0.02) |
| HD36841 | 14.2 (1.4) | 13.80 (0.04) | 13.80 (0.04) | 13.80 (0.06) |
| HD37021 | 30.1 (0.7) | 14.27 (0.01) | 14.34 (0.03) | 14.34 (0.03) |
| HD37903 | 16.8 (0.6) | 13.93 (0.02) | 13.96 (0.03) | 13.96 (0.04) |
| HD71634 | 27.3 (1.1) | 14.17 (0.02) | 14.19 (0.02) | 14.19 (0.03) |
| HD72754 | 33.8 (1.6) | 14.28 (0.03) | 14.32 (0.03) | 14.32 (0.04) |
| HD75309 | 43.5 (1.2) | 14.41 (0.02) | 14.42 (0.02) | 14.42 (0.03) |
| HD79186 | 54.6 (2.2) | 14.53 (0.03) | 14.60 (0.05) | 14.60 (0.06) |
| HD91824 | 51.3 (2.4) | 14.42 (0.02) | 14.43 (0.02) | 14.43 (0.03) |
| HD91983 | 53.7 (2.0) | 14.45 (0.02) | 14.49 (0.03) | 14.49 (0.04) |
| HD111934 | 68.0 (4.9) | 14.57 (0.04) | 14.60 (0.03) | 14.60 (0.05) |
| HD148594 | 22.3 (1.0) | 14.09 (0.02) | 14.13 (0.02) | 14.13 (0.03) |
| HD152590 | 50.2 (1.1) | 14.50 (0.01) | 14.51 (0.02) | 14.51 (0.02) |
| HD156110 | 9.1 (0.6) | 13.58 (0.03) | 13.59 (0.03) | 13.59 (0.04) |

Table 3—Continued

| Star | $W_\lambda$ | $\log_{10}(N)_A$ | $\log_{10}(N)_P$ | $\log_{10}(N)^{a}$ |
|---|---|---|---|---|
| HD157857 | 42.3 (1.6) | 14.45 (0.02) | 14.48 (0.03) | 14.48 (0.04) |
| HD175360 | 23.7 (0.9) | 14.07 (0.02) | 14.07 (0.09) | 14.07 (0.09) |
| HD185418 | 42.5 (1.1) | 14.46 (0.02) | 14.50 (0.03) | 14.50 (0.03) |
| HD198478 | 47.0 (1.4) | 14.50 (0.02) | 14.55 (0.03) | 14.55 (0.03) |
| HD198781 | 39.1 (1.9) | 14.26 (0.02) | 14.29 (0.03) | 14.29 (0.04) |
| HD201345 | 40.0 (1.3) | 14.28 (0.02) | 14.28 (0.06) | 14.28 (0.06) |
| HD203532 | 20.9 (0.7) | 14.04 (0.02) | 14.08 (0.02) | 14.08 (0.03) |
| HD206773 | 45.9 (1.4) | 14.45 (0.02) | 14.47 (0.04) | 14.47 (0.04) |
| HD208440 | 48.4 (1.6) | 14.45 (0.02) | 14.48 (0.07) | 14.48 (0.07) |
| HD212791 | 34.2 (1.7) | 14.20 (0.03) | 14.21 (0.01) | 14.21 (0.03) |
| HD220057 | 24.8 (0.9) | 14.16 (0.02) | 14.20 (0.03) | 14.20 (0.03) |
| | | *Manganese: Mn II $\lambda$1197* | | |
| HD1383 | 70.9 (7.0) | 13.72 (0.05) | 13.74 (0.04) | 13.74 (0.06) |
| HD12323 | 77.3 (5.1) | 13.68 (0.03) | 13.67 (0.08) | 13.67 (0.08) |
| HD13268 | 81.7 (5.6) | 13.72 (0.03) | 13.73 (0.02) | 13.73 (0.04) |
| HD14434 | 79.8 (7.9) | 13.71 (0.05) | 13.74 (0.06) | 13.74 (0.07) |
| HD27778 | 23.3 (1.7) | 13.17 (0.04) | 13.18 (0.05) | 13.18 (0.06) |
| HD36841 | 15.7 (2.4) | 12.98 (0.07) | 13.06 (0.07) | 13.06 (0.09) |
| HD37367 | 33.4 (1.4) | 13.49 (0.03) | 13.56 (0.03) | 13.56 (0.04) |
| HD37903 | 20.8 (1.1) | 13.20 (0.03) | 13.34 (0.05) | 13.34 (0.06) |
| HD43818 | 74.3 (3.1) | 13.82 (0.03) | 13.82 (0.04) | 13.82 (0.05) |
| HD52266 | 47.6 (2.6) | 13.55 (0.03) | 13.56 (0.02) | 13.56 (0.04) |
| HD63005 | 42.3 (2.3) | 13.44 (0.03) | 13.45 (0.05) | 13.45 (0.06) |
| HD71634 | 35.3 (3.4) | 13.46 (0.06) | 13.48 (0.03) | 13.48 (0.06) |
| HD75309 | 36.6 (1.6) | 13.44 (0.02) | 13.45 (0.04) | 13.45 (0.04) |
| HD79186 | 50.5 (4.0) | 13.60 (0.05) | 13.62 (0.05) | 13.62 (0.07) |
| HD91824 | 47.7 (2.7) | 13.53 (0.03) | 13.56 (0.05) | 13.56 (0.06) |
| HD91983 | 57.7 (3.4) | 13.60 (0.03) | 13.61 (0.05) | 13.61 (0.06) |
| HD111934 | 75.6 (5.2) | 13.76 (0.04) | 13.77 (0.03) | 13.77 (0.05) |
| HD116852 | 49.5 (5.2) | 13.52 (0.05) | 13.58 (0.05) | 13.58 (0.06) |

Table 3—Continued

| Star | $W_\lambda$ | $\log_{10}(N)_A$ | $\log_{10}(N)_P$ | $\log_{10}(N)^{\rm a}$ |
|---|---|---|---|---|
| HD122879 | 63.6 (3.5) | 13.67 (0.03) | 13.72 (0.03) | 13.72 (0.04) |
| HD147888 | 32.6 (1.1) | 13.52 (0.05) | 13.58 (0.05) | 13.58 (0.07) |
| HD148594 | 20.7 (2.2) | 13.21 (0.06) | 13.28 (0.07) | 13.28 (0.08) |
| HD152590 | 53.0 (2.7) | 13.68 (0.04) | 13.70 (0.08) | 13.70 (0.09) |
| HD157857 | 56.3 (4.7) | 13.69 (0.05) | 13.71 (0.06) | 13.71 (0.07) |
| HD165955 | 45.3 (2.3) | 13.46 (0.02) | 13.47 (0.03) | 13.47 (0.04) |
| HD175360 | 29.3 (2.2) | 13.37 (0.05) | 13.40 (0.07) | 13.40 (0.08) |
| HD185418 | 39.5 (1.7) | 13.59 (0.04) | 13.73 (0.04) | 13.73 (0.05) |
| HD190918 | 83.8 (4.1) | 13.75 (0.03) | 13.76 (0.05) | 13.76 (0.06) |
| HD192035 | 36.2 (2.9) | 13.35 (0.04) | 13.38 (0.06) | 13.38 (0.07) |
| HD192639 | 50.2 (2.8) | 13.67 (0.04) | 13.73 (0.07) | 13.73 (0.08) |
| HD198478 | 54.1 (3.1) | 13.71 (0.05) | 13.74 (0.08) | 13.74 (0.09) |
| HD198781 | 43.0 (3.2) | 13.42 (0.04) | 13.42 (0.03) | 13.42 (0.05) |
| HD201345 | 47.9 (2.1) | 13.52 (0.02) | 13.52 (0.02) | 13.52 (0.03) |
| HD203532 | 19.0 (1.7) | 13.15 (0.05) | 13.19 (0.03) | 13.19 (0.05) |
| HD206773 | 50.8 (3.2) | 13.63 (0.04) | 13.64 (0.04) | 13.64 (0.05) |
| HD207198 | 55.5 (2.5) | 13.69 (0.03) | 13.71 (0.06) | 13.71 (0.07) |
| HD208440 | 48.6 (3.4) | 13.55 (0.04) | 13.56 (0.05) | 13.56 (0.06) |
| HD210809 | 77.9 (3.8) | 13.71 (0.03) | 13.73 (0.03) | 13.73 (0.04) |
| HD212791 | 40.0 (2.1) | 13.42 (0.03) | 13.42 (0.02) | 13.42 (0.04) |
| HD220057 | 20.5 (1.5) | 13.16 (0.04) | 13.18 (0.05) | 13.18 (0.06) |
| HD232522 | 48.0 (3.8) | 13.55 (0.04) | 13.55 (0.04) | 13.55 (0.06) |
| HD308813 | 65.4 (3.9) | 13.61 (0.03) | 13.60 (0.02) | 13.60 (0.04) |
| BD+53 2820 | 86.2 (8.0) | 13.78 (0.05) | 13.85 (0.05) | 13.85 (0.06) |
| CPD−69 1743 | 45.7 (4.1) | 13.43 (0.04) | 13.47 (0.04) | 13.47 (0.05) |
| | | *Nickel: Ni II $\lambda$1317* | | |
| HD1383 | 71.5 (3.6) | 13.86 (0.02) | 13.87 (0.02) | 13.87 (0.03) |
| HD12323 | 74.6 (3.0) | 13.89 (0.02) | 13.89 (0.02) | 13.89 (0.03) |
| HD13268 | 90.1 (3.1) | 13.99 (0.02) | 14.00 (0.03) | 14.00 (0.04) |
| HD14434 | 98.9 (4.8) | 14.00 (0.02) | 14.01 (0.03) | 14.01 (0.04) |

Table 3—Continued

| Star | $W_\lambda$ | $\log_{10}(N)_A$ | $\log_{10}(N)_P$ | $\log_{10}(N)$[a] |
|---|---|---|---|---|
| HD27778 | 9.1 (0.5) | 12.93 (0.03) | 12.93 (0.03) | 12.93 (0.04) |
| HD36841 | 9.7 (1.4) | 12.93 (0.06) | 12.94 (0.05) | 12.94 (0.07) |
| HD37021 | 31.5 (0.7) | 13.65 (0.01) | 13.69 (0.02) | 13.69 (0.02) |
| HD37903 | 10.3 (0.5) | 13.00 (0.02) | 13.02 (0.02) | 13.02 (0.03) |
| HD43818 | 58.0 (1.0) | 13.84 (0.01) | 13.84 (0.03) | 13.84 (0.03) |
| HD52266 | 41.5 (1.8) | 13.64 (0.02) | 13.64 (0.01) | 13.64 (0.02) |
| HD63005 | 44.0 (1.1) | 13.71 (0.01) | 13.68 (0.02) | 13.68 (0.02) |
| HD71634 | 16.5 (1.1) | 13.20 (0.03) | 13.21 (0.06) | 13.21 (0.07) |
| HD75309 | 24.4 (1.2) | 13.38 (0.02) | 13.38 (0.05) | 13.38 (0.05) |
| HD79186 | 34.1 (0.7) | 13.57 (0.01) | 13.59 (0.02) | 13.59 (0.02) |
| HD91824 | 47.7 (1.7) | 13.71 (0.02) | 13.71 (0.05) | 13.71 (0.05) |
| HD91983 | 51.0 (1.5) | 13.74 (0.01) | 13.75 (0.05) | 13.75 (0.05) |
| HD111934 | 72.3 (3.2) | 13.85 (0.02) | 13.89 (0.11) | 13.89 (0.11) |
| HD116852 | 82.1 (4.0) | 13.93 (0.02) | 13.94 (0.03) | 13.44 (0.04) |
| HD122879 | 45.3 (1.2) | 13.64 (0.01) | 13.64 (0.01) | 13.64 (0.01) |
| HD147888 | 22.8 (0.5) | 13.48 (0.01) | 13.72 (0.06) | 13.72 (0.06) |
| HD148594 | 15.9 (0.9) | 13.23 (0.03) | 13.27 (0.02) | 13.27 (0.03) |
| HD157857 | 57.6 (1.9) | 13.81 (0.02) | 13.82 (0.03) | 13.82 (0.04) |
| HD165955 | 59.1 (2.0) | 13.79 (0.02) | 13.81 (0.08) | 13.81 (0.08) |
| HD175360 | 16.3 (0.7) | 13.20 (0.02) | 13.21 (0.05) | 13.21 (0.05) |
| HD185418 | 26.7 (1.0) | 13.46 (0.02) | 13.48 (0.02) | 13.48 (0.03) |
| HD190918 | 86.3 (1.9) | 13.98 (0.01) | 13.98 (0.03) | 13.98 (0.03) |
| HD192035 | 45.3 (1.8) | 13.60 (0.02) | 13.61 (0.06) | 13.61 (0.06) |
| HD192639 | 49.8 (1.6) | 13.81 (0.02) | 13.80 (0.06) | 13.80 (0.06) |
| HD198478 | 37.5 (1.2) | 13.62 (0.02) | 13.63 (0.07) | 13.63 (0.07) |
| HD198781 | 43.3 (1.3) | 13.63 (0.01) | 13.64 (0.02) | 13.64 (0.02) |
| HD201345 | 61.9 (1.3) | 13.86 (0.01) | 13.87 (0.04) | 13.87 (0.04) |
| HD203532 | 7.7 (0.8) | 12.84 (0.05) | 12.84 (0.06) | 12.84 (0.08) |
| HD206773 | 34.9 (1.6) | 13.59 (0.02) | 13.60 (0.05) | 13.60 (0.05) |
| HD207198 | 45.5 (0.6) | 13.72 (0.01) | 13.73 (0.06) | 13.73 (0.06) |

Table 3—Continued

| Star | $W_\lambda$ | $\log_{10}(N)_A$ | $\log_{10}(N)_P$ | $\log_{10}(N)$[a] |
|---|---|---|---|---|
| HD208440 | 37.4 (1.1) | 13.61 (0.02) | 13.61 (0.05) | 13.61 (0.05) |
| HD210809 | 97.7 (2.4) | 14.03 (0.01) | 14.03 (0.03) | 14.03 (0.03) |
| HD212791 | 43.7 (1.3) | 13.65 (0.01) | 13.66 (0.08) | 13.66 (0.08) |
| HD220057 | 21.8 (1.1) | 13.37 (0.02) | 13.38 (0.05) | 13.38 (0.05) |
| HD232522 | 55.8 (1.5) | 13.84 (0.01) | 13.84 (0.04) | 13.84 (0.04) |
| HD308813 | 68.6 (2.3) | 13.85 (0.02) | 13.85 (0.05) | 13.85 (0.05) |
| BD+53 2820 | 112.8 (6.1) | 14.09 (0.03) | 14.09 (0.03) | 14.09 (0.04) |
| CPD−69 1743 | 68.4 (2.7) | 13.84 (0.02) | 13.86 (0.03) | 13.86 (0.04) |
| | | *Copper: Cu II $\lambda$1358* | | |
| HD1383 | 21.3 (3.7) | 12.58 (0.07) | 12.57 (0.04) | 12.57 (0.08) |
| HD12323 | 13.7 (1.3) | 12.37 (0.04) | 12.39 (0.04) | 12.39 (0.05) |
| HD13268 | 18.0 (2.0) | 12.49 (0.05) | 12.51 (0.04) | 12.51 (0.05) |
| HD14434 | 26.4 (3.5) | 12.66 (0.06) | 12.67 (0.04) | 12.67 (0.07) |
| HD27778 | 7.0 (0.4) | 12.08 (0.03) | 12.10 (0.04) | 12.10 (0.05) |
| HD36841 | 4.8 (0.9) | 11.90 (0.07) | 11.89 (0.07) | 11.89 (0.10) |
| HD37021 | 19.0 (0.7) | 12.61 (0.02) | 12.65 (0.02) | 12.65 (0.03) |
| HD37061 | 20.7 (0.6) | 12.71 (0.01) | 12.79 (0.02) | 12.79 (0.02) |
| HD37367 | 13.8 (0.6) | 12.41 (0.02) | 12.43 (0.05) | 12.43 (0.05) |
| HD37903 | 8.0 (0.6) | 12.16 (0.03) | 12.19 (0.03) | 12.19 (0.04) |
| HD43818 | 29.4 (1.6) | 12.74 (0.02) | 12.73 (0.02) | 12.73 (0.02) |
| HD52266 | 15.8 (1.4) | 12.44 (0.04) | 12.45 (0.02) | 12.45 (0.03) |
| HD63005 | 13.6 (1.2) | 12.37 (0.04) | 12.38 (0.03) | 12.38 (0.05) |
| HD71634 | 9.5 (0.8) | 12.22 (0.04) | 12.24 (0.03) | 12.24 (0.05) |
| HD75309 | 11.9 (1.2) | 12.31 (0.04) | 12.32 (0.04) | 12.32 (0.05) |
| HD79186 | 15.0 (1.0) | 12.43 (0.03) | 12.47 (0.03) | 12.47 (0.04) |
| HD91824 | 12.3 (1.7) | 12.32 (0.06) | 12.33 (0.03) | 12.33 (0.07) |
| HD91983 | 16.6 (1.9) | 12.46 (0.05) | 12.43 (0.08) | 12.43 (0.09) |
| HD111934 | 26.1 (3.0) | 12.67 (0.05) | 12.66 (0.02) | 12.66 (0.05) |
| HD116852 | 7.6 (1.2) | 12.12 (0.06) | 12.10 (0.05) | 12.10 (0.08) |
| HD122879 | 19.5 (2.0) | 12.53 (0.04) | 12.54 (0.04) | 12.54 (0.05) |

Table 3—Continued

| Star | $W_\lambda$ | $\log_{10}(N)_A$ | $\log_{10}(N)_P$ | $\log_{10}(N)$[a] |
|---|---|---|---|---|
| HD148594 | 8.8 (0.6) | 12.22 (0.03) | 12.26 (0.02) | 12.26 (0.03) |
| HD152590 | 18.3 (1.3) | 12.53 (0.03) | 12.55 (0.08) | 12.55 (0.08) |
| HD157857 | 18.0 (1.0) | 12.54 (0.02) | 12.54 (0.03) | 12.54 (0.04) |
| HD165955 | 16.2 (1.4) | 12.44 (0.04) | 12.42 (0.05) | 12.42 (0.06) |
| HD175360 | 6.1 (0.9) | 12.02 (0.06) | 12.05 (0.07) | 12.05 (0.09) |
| HD185418 | 16.1 (1.1) | 12.48 (0.03) | 12.47 (0.03) | 12.47 (0.04) |
| HD190918 | 23.6 (2.0) | 12.61 (0.04) | 12.63 (0.08) | 12.63 (0.09) |
| HD192035 | 15.9 (1.3) | 12.44 (0.03) | 12.47 (0.08) | 12.47 (0.08) |
| HD192639 | 22.4 (1.1) | 12.64 (0.02) | 12.65 (0.04) | 12.65 (0.04) |
| HD198478 | 20.3 (0.9) | 12.58 (0.02) | 12.59 (0.02) | 12.59 (0.03) |
| HD201345 | 13.9 (1.6) | 12.38 (0.05) | 12.35 (0.04) | 12.35 (0.07) |
| HD203532 | 7.5 (0.6) | 12.12 (0.03) | 12.13 (0.06) | 12.13 (0.07) |
| HD208440 | 18.4 (1.9) | 12.51 (0.04) | 12.51 (0.02) | 12.51 (0.04) |
| HD210809 | 19.1 (1.9) | 12.51 (0.04) | 12.51 (0.08) | 12.51 (0.09) |
| HD212791 | 9.2 (1.2) | 12.19 (0.05) | 12.19 (0.05) | 12.19 (0.07) |
| HD220057 | 9.0 (0.8) | 12.20 (0.04) | 12.23 (0.06) | 12.23 (0.07) |
| HD308813 | 20.6 (1.9) | 12.55 (0.04) | 12.55 (0.03) | 12.55 (0.05) |
| BD+53 2820 | 15.2 (3.1) | 12.41 (0.08) | 12.44 (0.06) | 12.44 (0.09) |
| CPD−69 1743 | 17.7 (2.3) | 12.48 (0.05) | 12.50 (0.04) | 12.50 (0.06) |
| | | *Germanium: Ge II* $\lambda$*1237* | | |
| HD1383 | 38.7 (4.5) | 12.45 (0.05) | 12.45 (0.08) | 12.45 (0.09) |
| HD12323 | 20.4 (1.7) | 12.14 (0.04) | 12.15 (0.05) | 12.15 (0.06) |
| HD13268 | 39.3 (5.3) | 12.43 (0.06) | 12.44 (0.04) | 12.44 (0.07) |
| HD27778 | 17.4 (0.7) | 12.12 (0.02) | 12.14 (0.03) | 12.14 (0.04) |
| HD37021 | 23.1 (1.2) | 12.29 (0.03) | 12.32 (0.03) | 12.32 (0.04) |
| HD37061 | 24.5 (0.8) | 12.41 (0.02) | 12.51 (0.05) | 12.51 (0.05) |
| HD37367 | 24.7 (1.0) | 12.32 (0.02) | 12.34 (0.03) | 12.34 (0.04) |
| HD37903 | 13.0 (0.6) | 11.98 (0.02) | 12.02 (0.03) | 12.02 (0.03) |
| HD43818 | 44.0 (4.3) | 12.57 (0.05) | 12.58 (0.03) | 12.58 (0.06) |
| HD52266 | 34.2 (2.1) | 12.43 (0.02) | 12.43 (0.02) | 12.43 (0.04) |

Table 3—Continued

| Star | $W_\lambda$ | $\log_{10}(N)_A$ | $\log_{10}(N)_P$ | $\log_{10}(N)$[a] |
|---|---|---|---|---|
| HD63005 | 26.8 (1.7) | 12.29 (0.03) | 12.31 (0.03) | 12.31 (0.04) |
| HD71634 | 13.5 (0.9) | 11.99 (0.03) | 12.02 (0.05) | 12.02 (0.06) |
| HD72754 | 20.2 (1.2) | 12.17 (0.03) | 12.20 (0.04) | 12.20 (0.05) |
| HD75309 | 22.4 (1.4) | 12.21 (0.03) | 12.22 (0.05) | 12.22 (0.06) |
| HD79186 | 28.0 (1.5) | 12.33 (0.03) | 12.34 (0.02) | 12.34 (0.06) |
| HD91824 | 29.8 (3.9) | 12.32 (0.06) | 12.33 (0.07) | 12.33 (0.08) |
| HD91983 | 32.4 (2.1) | 12.37 (0.03) | 12.37 (0.02) | 12.37 (0.04) |
| HD111934 | 47.0 (3.0) | 12.55 (0.03) | 12.56 (0.03) | 12.56 (0.04) |
| HD116852 | 13.1 (1.7) | 11.97 (0.05) | 12.00 (0.05) | 12.00 (0.07) |
| HD122879 | 38.5 (4.5) | 12.45 (0.05) | 12.47 (0.06) | 12.47 (0.07) |
| HD147888 | 19.7 (0.7) | 12.27 (0.02) | 12.48 (0.06) | 12.48 (0.06) |
| HD148594 | 15.6 (1.5) | 12.10 (0.04) | 12.15 (0.05) | 12.15 (0.06) |
| HD152590 | 32.9 (1.4) | 12.42 (0.02) | 12.44 (0.05) | 12.44 (0.05) |
| HD156110 | 3.6 (0.6) | 11.36 (0.06) | 11.37 (0.03) | 11.37 (0.07) |
| HD157857 | 38.3 (3.2) | 12.54 (0.05) | 12.52 (0.07) | 12.52 (0.09) |
| HD165955 | 23.9 (1.1) | 12.20 (0.02) | 12.19 (0.03) | 12.19 (0.04) |
| HD175360 | 11.8 (1.2) | 11.90 (0.04) | 11.91 (0.05) | 11.91 (0.06) |
| HD185418 | 29.7 (1.4) | 12.41 (0.02) | 12.44 (0.03) | 12.44 (0.04) |
| HD192035 | 29.2 (1.8) | 12.31 (0.03) | 12.33 (0.04) | 12.33 (0.05) |
| HD198478 | 36.5 (1.6) | 12.48 (0.02) | 12.48 (0.04) | 12.48 (0.04) |
| HD198781 | 22.7 (2.5) | 12.19 (0.05) | 12.20 (0.03) | 12.20 (0.06) |
| HD201345 | 18.2 (2.0) | 12.09 (0.05) | 12.09 (0.03) | 12.09 (0.06) |
| HD203532 | 14.5 (0.9) | 12.03 (0.03) | 12.07 (0.04) | 12.07 (0.05) |
| HD206773 | 28.1 (1.7) | 12.33 (0.03) | 12.34 (0.04) | 12.34 (0.05) |
| HD207198 | 35.1 (1.3) | 12.47 (0.02) | 12.48 (0.05) | 12.48 (0.05) |
| HD208440 | 29.7 (1.9) | 12.34 (0.03) | 12.35 (0.05) | 12.35 (0.06) |
| HD210809 | 51.6 (6.5) | 12.56 (0.06) | 12.56 (0.02) | 12.56 (0.06) |
| HD212791 | 21.7 (1.6) | 12.16 (0.03) | 12.15 (0.06) | 12.15 (0.07) |
| HD220057 | 17.4 (1.1) | 12.13 (0.03) | 12.15 (0.04) | 12.15 (0.05) |
| HD232522 | 24.1 (2.7) | 12.25 (0.05) | 12.25 (0.06) | 12.25 (0.08) |

Table 3—Continued

| Star | $W_\lambda$ | $\log_{10}(N)_A$ | $\log_{10}(N)_P$ | $\log_{10}(N)^a$ |
|---|---|---|---|---|
| HD308813 | 35.1 (2.6) | 12.37 (0.03) | 12.37 (0.04) | 12.37 (0.05) |
| BD+53 2820 | 32.9 (6.2) | 12.34 (0.08) | 12.38 (0.05) | 12.38 (0.09) |
| CPD−69 1743 | 25.2 (3.3) | 12.22 (0.06) | 12.24 (0.03) | 12.24 (0.06) |

[a]The adopted column densities for each element are profile-fit results. The uncertainties assigned are the quadrature sum of absolute errors for statistical error, derived from the apparent-optical-depth method, and profile-fitting uncertainties.

[b]The equivalent widths quoted for the magnesium doublet refer only to the stronger line at 1239.9253 Å.

Note. — Measurements are tabulated for the apparent-optical-depth (designated by the subscript $A$) and profile-fitting (designated by subscript $P$) methods of deriving elemental column densities. The values adopted for further analysis are presented in the rightmost column.

Table 4. Elemental Abundances

| Star | $\log_{10}[N(\mathrm{H})]$ | $\log_{10}\langle n_{\mathrm{H}}\rangle$ | $\log_{10}(\mathrm{Mg/H})$ | $\log_{10}(\mathrm{P/H})$ | $\log_{10}(\mathrm{Mn/H})$ | $\log_{10}(\mathrm{Ni/H})$ | $\log_{10}(\mathrm{Cu/H})$ | $\log_{10}(\mathrm{Ge/H})$ |
|---|---|---|---|---|---|---|---|---|
| HD1383 | 21.50 (0.08) | -0.45 | -5.14 (0.12) | ... | -7.76 (0.10) | -7.63 (0.08) | -8.93 (0.11) | -9.05 (0.12) |
| HD12323 | 21.29 (0.07) | -0.56 | -5.25 (0.08) | ... | -7.62 (0.10) | -7.40 (0.08) | -8.90 (0.08) | -9.14 (0.09) |
| HD13268 | 21.42 (0.07) | -0.43 | -5.18 (0.08) | ... | -7.69 (0.08) | -7.42 (0.08) | -8.91 (0.08) | -8.98 (0.10) |
| HD14434 | 21.47 (0.07) | -0.38 | -5.20 (0.08) | ... | -7.73 (0.10) | -7.46 (0.08) | -8.80 (0.10) | ... |
| HD27778 | 21.36 (0.08) | 0.53 | -5.88 (0.08) | -7.28 (0.08) | -8.18 (0.10) | -8.43 (0.09) | -9.26 (0.09) | -9.22 (0.09) |
| HD36841[a] | 21.18 (0.12) | 0.13 | -5.73 (0.13) | -7.38 (0.13) | -8.12 (0.14) | -8.24 (0.13) | -9.29 (0.15) | ... |
| HD37021 | 21.68 (0.12) | 0.61 | -5.78 (0.12) | -7.34 (0.12) | ... | -7.99 (0.12) | -9.03 (0.12) | -9.36 (0.12) |
| HD37061 | 21.73 (0.09) | 0.66 | -5.93 (0.10) | ... | ... | ... | -8.94 (0.09) | -9.22 (0.10) |
| HD37367 | 21.41 (0.07) | 0.36 | -5.41 (0.08) | ... | -7.85 (0.08) | ... | -8.98 (0.08) | -9.07 (0.08) |
| HD37903 | 21.46 (0.06) | 0.41 | -5.84 (0.08) | -7.50 (0.07) | -8.12 (0.08) | -8.44 (0.07) | -9.27 (0.07) | -9.44 (0.07) |
| HD43818[a] | 21.71 (0.11) | 0.04 | -5.23 (0.12) | ... | -7.89 (0.12) | -7.87 (0.11) | -8.98 (0.11) | -9.13 (0.12) |
| HD52266[a] | 21.42 (0.10) | -0.31 | -5.33 (0.10) | ... | -7.86 (0.11) | -7.78 (0.10) | -8.97 (0.10) | -8.99 (0.11) |
| HD63005 | 21.32 (0.05) | -0.57 | -5.29 (0.06) | ... | -7.87 (0.08) | -7.64 (0.05) | -8.94 (0.07) | -9.01 (0.06) |
| HD71634[a] | 20.90 (0.11) | -0.19 | -5.11 (0.12) | -6.71 (0.11) | -7.42 (0.12) | -7.69 (0.13) | -8.66 (0.12) | -8.88 (0.12) |
| HD72754 | 21.29 (0.10) | -0.04 | ... | -6.97 (0.11) | ... | ... | ... | -9.09 (0.11) |
| HD75309 | 21.18 (0.08) | -0.55 | -5.23 (0.09) | -6.76 (0.08) | -7.73 (0.09) | -7.80 (0.09) | -8.86 (0.09) | -8.96 (0.10) |
| HD79186 | 21.42 (0.07) | -0.31 | -5.31 (0.08) | -6.82 (0.09) | -7.80 (0.10) | -7.83 (0.07) | -8.95 (0.08) | -9.08 (0.09) |
| HD91824 | 21.16 (0.05) | -0.73 | -5.02 (0.06) | -6.73 (0.06) | -7.60 (0.08) | -7.45 (0.07) | -8.83 (0.08) | -8.83 (0.09) |
| HD91983 | 21.24 (0.08) | -0.65 | -5.05 (0.10) | -6.75 (0.09) | -7.63 (0.10) | -7.49 (0.09) | -8.81 (0.12) | -8.87 (0.09) |
| HD111934[a] | 21.58 (0.12) | -0.29 | -5.30 (0.12) | -6.98 (0.13) | -7.81 (0.13) | -7.69 (0.16) | -8.92 (0.13) | -9.02 (0.12) |
| HD116852 | 21.02 (0.08) | -1.15 | -5.11 (0.09) | ... | -7.44 (0.10) | -7.58 (0.09) | -8.92 (0.11) | -9.02 (0.10) |
| HD122879 | 21.34 (0.10) | -0.47 | -5.11 (0.10) | ... | -7.62 (0.11) | -7.70 (0.10) | -8.80 (0.11) | -8.87 (0.12) |

Table 4—Continued

| Star | $\log_{10}[N(\mathrm{H})]$ | $\log_{10}\langle n_{\mathrm{H}}\rangle$ | $\log_{10}(\mathrm{Mg/H})$ | $\log_{10}(\mathrm{P/H})$ | $\log_{10}(\mathrm{Mn/H})$ | $\log_{10}(\mathrm{Ni/H})$ | $\log_{10}(\mathrm{Cu/H})$ | $\log_{10}(\mathrm{Ge/H})$ |
|---|---|---|---|---|---|---|---|---|
| HD147888 | 21.76 (0.08) | 1.10 | -5.73 (0.08) | ... | -8.18 (0.10) | -8.04 (0.10) | ... | -9.28 (0.10) |
| HD148594[a] | 21.39 (0.10) | 0.72 | -5.78 (0.11) | -7.26 (0.10) | -8.11 (0.12) | -8.12 (0.10) | -9.13 (0.10) | -9.24 (0.11) |
| HD152590 | 21.47 (0.05) | -0.28 | -5.27 (0.10) | -6.96 (0.05) | -7.77 (0.10) | ... | -8.92 (0.09) | -9.03 (0.07) |
| HD156110 | 20.66 (0.13) | -0.69 | -5.52 (0.13) | -7.07 (0.13) | ... | ... | ... | -9.29 (0.14) |
| HD157857 | 21.44 (0.07) | -0.41 | -5.25 (0.09) | -6.96 (0.08) | -7.73 (0.10) | -7.62 (0.08) | -8.90 (0.08) | -8.92 (0.11) |
| HD165955 | 21.11 (0.06) | -0.59 | -5.09 (0.08) | ... | -7.64 (0.07) | -7.30 (0.10) | -8.69 (0.08) | -8.92 (0.07) |
| HD175360[a] | 21.03 (0.11) | 0.11 | -5.43 (0.12) | -6.96 (0.14) | -7.63 (0.13) | -7.82 (0.12) | -8.98 (0.14) | -9.12 (0.12) |
| HD185418 | 21.41 (0.07) | 0.08 | -5.45 (0.07) | -6.91 (0.08) | -7.68 (0.08) | -7.93 (0.08) | -8.94 (0.08) | -8.97 (0.08) |
| HD190918 | 21.40 (0.06) | -0.45 | -5.06 (0.07) | ... | -7.64 (0.08) | -7.42 (0.07) | -8.77 (0.11) | ... |
| HD192035 | 21.38 (0.07) | -0.50 | -5.45 (0.08) | ... | -8.00 (0.10) | -7.77 (0.09) | -8.91 (0.10) | -9.05 (0.08) |
| HD192639 | 21.48 (0.07) | -0.27 | -5.27 (0.08) | ... | -7.75 (0.10) | -7.68 (0.09) | -8.83 (0.08) | ... |
| HD198478 | 21.55 (0.11) | 0.16 | -5.60 (0.12) | -7.00 (0.11) | -7.81 (0.14) | -7.92 (0.13) | -8.96 (0.11) | -9.07 (0.12) |
| HD198781 | 21.15 (0.06) | -0.13 | -5.39 (0.07) | -6.86 (0.07) | -7.73 (0.08) | -7.51 (0.06) | ... | -8.95 (0.08) |
| HD201345 | 21.00 (0.08) | -0.90 | -5.04 (0.08) | -6.72 (0.10) | -7.48 (0.08) | -7.13 (0.09) | -8.65 (0.10) | -8.91 (0.10) |
| HD203532 | 21.44 (0.07) | 0.55 | -5.86 (0.07) | -7.36 (0.08) | -8.25 (0.08) | -8.60 (0.10) | -9.31 (0.10) | -9.37 (0.08) |
| HD206773 | 21.25 (0.05) | -0.03 | -5.26 (0.05) | -6.78 (0.06) | -7.61 (0.07) | -7.65 (0.07) | ... | -8.91 (0.07) |
| HD207198 | 21.68 (0.09) | 0.40 | -5.60 (0.09) | ... | -7.97 (0.11) | -7.95 (0.11) | ... | -9.20 (0.10) |
| HD208440 | 21.32 (0.08) | 0.04 | -5.27 (0.09) | -6.84 (0.10) | -7.76 (0.10) | -7.71 (0.09) | -8.81 (0.09) | -8.97 (0.10) |
| HD210809 | 21.33 (0.08) | -0.70 | -5.10 (0.10) | ... | -7.60 (0.09) | -7.30 (0.08) | -8.82 (0.12) | -8.77 (0.10) |
| HD212791 | 21.22 (0.09) | 0.16 | -5.35 (0.11) | -7.01 (0.09) | -7.80 (0.10) | -7.56 (0.12) | -9.03 (0.11) | -9.07 (0.11) |
| HD220057 | 21.27 (0.07) | 0.14 | -5.61 (0.08) | -7.07 (0.08) | -8.09 (0.09) | -7.89 (0.08) | -9.04 (0.10) | -9.12 (0.08) |
| HD232522 | 21.19 (0.05) | -0.92 | -5.08 (0.06) | ... | -7.64 (0.08) | -7.35 (0.06) | ... | -8.94 (0.09) |

Table 4—Continued

| Star | $\log_{10}[N(\mathrm{H})]$ | $\log_{10}\langle n_{\mathrm{H}}\rangle$ | $\log_{10}$(Mg/H) | $\log_{10}$(P/H) | $\log_{10}$(Mn/H) | $\log_{10}$(Ni/H) | $\log_{10}$(Cu/H) | $\log_{10}$(Ge/H) |
|---|---|---|---|---|---|---|---|---|
| HD308813 | 21.29 (0.08) | -0.18 | -5.14 (0.08) | ... | -7.69 (0.09) | -7.44 (0.09) | -8.74 (0.09) | -8.92 (0.09) |
| BD +53 2820 | 21.39 (0.08) | -0.64 | ... | ... | -7.54 (0.11) | -7.30 (0.10) | -8.95 (0.12) | -9.01 (0.12) |
| CPD −69 1743 | 21.16 (0.08) | -1.15 | ... | ... | -7.69 (0.09) | -7.30 (0.09) | -8.66 (0.10) | -8.92 (0.10) |

[a]The hydrogen column densities for the sight lines toward HD36841, HD43818, HD52266, HD71634, HD111934, HD148594, and HD175360 were assigned by calibration to the krypton abundance or, in its absence, the oxygen abundance.

Table 5. Boltzmann Parameters: Lightly-Depleted Elements

| Element | $(x/\mathrm{H})_w$ | $(x/\mathrm{H})_c$ | $\langle n_0 \rangle$ (cm$^{-3}$) | $m$ (cm$^{-3}$) | $\chi^2_\nu$ | Scatter (dex) |
|---|---|---|---|---|---|---|
| Phosphorus | $1.36{\pm}0.08{\times}10^{-7}$ | $3.97{\pm}0.39{\times}10^{-08}$ | 1.53±0.09 | 0.199±0.145 | 1.24 | 0.113 |
| Germanium | $1.04{\pm}0.03{\times}10^{-9}$ | $5.23{\pm}0.23{\times}10^{-10}$ | 1.44±0.11 | 0.199±0.131 | 0.91 | 0.097 |
| Oxygen | $3.90{\pm}0.10{\times}10^{-4}$ | $2.84{\pm}0.12{\times}10^{-04}$ | 1.50±0.22 | 0.231±0.245 | 0.69 | 0.089 |
| Krypton | $9.63{\pm}0.24{\times}10^{-10}$ | | ··· | ··· | ··· | 0.062 |

Note. — Krypton is included for comparison as a lightly depleted element, although abundance data for this element was fit with a straight line as a function of mean hydrogen sight line density rather than the Boltzmann function defined in § 3.

Table 6. Boltzmann Parameters: Mg, Mn, Cu, Ni

| Element | $(x/\mathrm{H})_w$ | $(x/\mathrm{H})_d$ | $(x/\mathrm{H})_c$ | $\langle n_0\rangle$ ($\mathrm{cm}^{-3}$) | $\langle n_1\rangle$ ($\mathrm{cm}^{-3}$) | $m_0$ ($\mathrm{cm}^{-3}$) | $m_1$ ($\mathrm{cm}^{-3}$) | $\chi^2_\nu$ | Scatter (dex) |
|---|---|---|---|---|---|---|---|---|---|
| | | | | *Two Plateau Fits* | | | | | |
| Magnesium | $1.60\pm0.06\times10^{-5}$ | ··· | $1.48\pm0.07\times10^{-6}$ | $-0.13\pm0.04$ | ··· | $0.736\pm0.036$ | ··· | 1.33 | 0.122 |
| Manganese | $3.71\pm1.13\times10^{-8}$ | ··· | $5.86\pm1.07\times10^{-9}$ | $0.29\pm0.65$ | ··· | $0.934\pm0.259$ | ··· | 1.21 | 0.134 |
| Copper | $1.35\pm0.15\times10^{-9}$ | ··· | $6.72\pm0.51\times10^{-10}$ | $1.17\pm0.27$ | ··· | $0.317\pm0.301$ | ··· | 1.23 | 0.132 |
| Nickel | $8.35\pm13.4\times10^{-6}$ | ··· | $1.06\pm0.07\times10^{-8}$ | $-2.30\pm0.71$ | ··· | $0.424\pm0.069$ | ··· | 1.98 | 0.169 |
| | | | | *Three Plateau Fits*[a] | | | | | |
| Magnesium | $9.05\pm0.63\times10^{-6}$ | $5.96\pm0.22\times10^{-6}$ | $1.59\pm0.07\times10^{-6}$ | $0.22\pm0.02$ | $1.18\pm0.06$ | $0.017\pm0.017$ | $0.351\pm0.062$ | 1.49 | 0.135 |
| Nickel | $4.43\pm0.47\times10^{-8}$ | $2.12\pm0.14\times10^{-8}$ | $1.05\pm0.08\times10^{-8}$ | $0.24\pm0.02$ | $1.13\pm0.09$ | $0.012\pm0.013$ | $0.088\pm0.086$ | 1.82 | 0.175 |

Note. — In order to derive stable solutions for the Boltzmann parameters that accurately reflected the data, individual sight lines for each element were pruned from the corresponding dataset; these paths, however, were re-introduced into each dataset before the scatter was determined. See § 3.2 for the factors likely to degrade the fits for heavily depleted elements. For magnesium, HD37367, HD156110, and HD192035 were excluded; for manganese, HD63005, HD116852, HD192035, and $\zeta$ Oph; for copper, $\kappa$ Ori; for nickel, HD27778, HD36841, HD37903, HD75309, HD116852, HD192035, and HD203532.

[a]The three-plateau Boltzmann function is defined by: $(x/\mathrm{H})_{gas} = (x/\mathrm{H})_c + \frac{(x/\mathrm{H})_w - (x/\mathrm{H})_d}{1+e^{(\langle n_\mathrm{H}\rangle - \langle n_1\rangle)/m_1}} + \frac{(x/\mathrm{H})_d - (x/\mathrm{H})_c}{1+e^{(\langle n_\mathrm{H}\rangle - \langle n_0\rangle)/m_0}}$, where $d$ refers to the cold diffuse ISM plateau intermediate between the $w$ warm and $c$ cold neutral medium plateaus.

Table 7. Fluctuations with Heliocentric Distance

| Element | Local Sample | Scatter (dex) | Distant Sample | Scatter (dex) | Gap |
|---|---|---|---|---|---|
| Copper | -8.992±0.027 | 0.164 | -8.863±0.020 | 0.102 | 0.129±0.033 |
| Oxygen | -3.461±0.022 | 0.060 | -3.362±0.018 | 0.076 | 0.099±0.028 |

Table 8. ISM Cloud Depletions

| | Solar | Warm ISM | | Cold ISM | |
|---|---|---|---|---|---|
| Element | Abundance | Previous[a] | Current Sample[b] | Previous[a] | Current Sample[b] |
| Oxygen | 8.69 | -0.22 | -0.10 | -0.22 | -0.24 |
| Magnesium | 7.55 | -0.57 | -0.59 | -1.17 | -1.38 |
| Phosphorus | 5.46 | -0.09 | -0.33 | -0.39 | -0.86 |
| Manganese | 5.50 | -0.97 | -0.93 | -1.47 | -1.73 |
| Nickel | 6.22 | -1.37 | -1.57 | -2.17 | -2.19 |
| Copper | 4.26 | -1.09 | -1.13 | -1.39 | -1.43 |
| Germanium | 3.62 | -0.49 | -0.60 | -0.79 | -0.90 |
| Krypton | 3.28 | -0.25 | -0.30 | -0.25 | -0.30 |

[a]The set of previous depletions are adopted from Welty et al. (1999), with adjustments for changes in the solar (cosmic) abundance standard (Lodders 2003). Welty et al. listed these levels to one decimal place; consequently, the second decimal place should be regarded as uncertain up to at least 0.05 dex.

[b]The set of depletions established from the current dataset are different from those in Tables 5 and 6 in the case of magnesium and nickel. For these elements, the adopted warm ISM depletion levels are those derived from the three-plateau Boltzmann solution.

Table 9. Dust Allowances per Million H Atoms

| | B STARS | | YOUNG F AND G STARS | | SUN (photosphere) | | SUN (protostellar) | |
|---|---|---|---|---|---|---|---|---|
| ELEMENT | Abundance[a] | High-$\langle n_{\rm H} \rangle$ | Abundance[a] | High-$\langle n_{\rm H} \rangle$ | Abundance[a] | High-$\langle n_{\rm H} \rangle$ | Abundance[a] | High-$\langle n_{\rm H} \rangle$ |
| Oxygen | 350 | 66 | 445 | 161 | 490 | 206 | 575 | 291 |
| Magnesium | 23.0 | 21.5 | 42.7 | 41.2 | 35.5 | 34.0 | 41.7 | 40.2 |
| Silicon[b] | 18.8 | 16.6 | 39.9 | 37.7 | 34.7 | 32.5 | 40.7 | 38.5 |
| Iron[b] | 28.5 | 28.2 | 27.9 | 27.6 | 29.5 | 29.2 | 34.7 | 34.3 |
| Predicted O[c] | ··· | 99 | ··· | 173 | ··· | 154 | ··· | 182 |

[a]Total elemental abundances in parts per million: for B stars and young F and G stars from Sofia & Meyer (2001), for the Sun from Lodders (2003).

[b]The weighted mean of Miller et al. (2005) translucent sight line silicon and iron abundances are $({\rm Si/H})_{gas} = (2.24\pm0.29)\times10^{-6}$ and $({\rm Fe/H})_{gas} = (0.26\pm0.02)\times10^{-6}$.

[c]The oxygen dust abundance predicted by observed magnesium, iron, and silicon gas-phase abundances, the various cosmic standards, and rudimentary dust models.